\documentclass[preprint,showpacs,showkeys, floatfix]{revtex4}
\usepackage{graphicx,amsmath,epsfig,amssymb,subfigure}

\newcommand{\sech}{\operatorname{sech}}
\newcommand{\diag}{\operatorname{diag}}

\begin{document}

\title{Vacuum Energy Density for Massless Scalar Fields in Flat
Homogeneous Spacetime Manifolds with Nontrivial Topology}

\author{P.~M.~Sutter} \email{psutter2@uiuc.edu}
\altaffiliation{\emph{Present address:} Department of Physics,
University of Illinois at Urbana-Champaign, 1110 West Green Street,
Urbana, IL 61801-3080}

\author{Tsunefumi Tanaka} \email{tt22@humboldt.edu}
\altaffiliation{\emph{Present address:} Department of Physics and
Physical Science, Humboldt State University, 1 Harpst Street, Arcata,
CA 95521} \affiliation{Physics Department, California Polytechnic
State University, San Luis Obispo, CA 93407}

\begin{abstract}
Although the observed universe appears to be geometrically flat, it
could have one of 18 global topologies.  A constant-time slice of the
spacetime manifold could be a torus, M\"{o}bius strip, Klein bottle,
or others.  This global topology of the universe imposes boundary
conditions on quantum fields and affects the vacuum energy density via
Casimir effect.  In a spacetime with such a nontrivial topology, the
vacuum energy density is shifted from its value in a simply-connected
spacetime.  In this paper, the vacuum expectation value of the
stress-energy tensor for a massless scalar field is calculated in all
17 multiply-connected, flat and homogeneous spacetimes with different
global topologies.  It is found that the vacuum energy density is
lowered relative to the Minkowski vacuum level in all spacetimes and
that the stress-energy tensor becomes position-dependent in spacetimes
that involve reflections and rotations.
\end{abstract}


\maketitle

\section{Introduction}
No known laws of physics can predict the global topology of spacetime.
However, the global topology of the universe plays an important role
in quantum field theory.  The geometrically flat, homogeneous, and
isotropic universe could have one of 18 possible spatial manifolds
with different topologies.  Only one of these, the Euclidean plane, is
simply-connected.  The other 17 are multiply-connected; that is, the
spacetime contains multiple geodesics between two points that cannot
be continuously deformed into one another.

If the observable part of the universe is greater in size than the
greatest simply-connected domain of a spacetime manifold, then the
global topology of the universe will be directly observable.  There
are currently two major efforts to observationally determine the
cosmic topology: cosmic crystallography and searches for circles in
the sky.  Researchers in cosmic crystallography use a statistical
approach; they examine the separations of random pairs of galaxies in
the sky.  If the universe is multiply-connected, certain distances
will appear more frequently in a statistical distribution.  This
method cannot determine the exact topology.  For example, it cannot
distinguish between a M\"{o}bius strip and a 1-Torus with twice the
circumference.

The second effort, the circles in the sky, assumes that the visible
universe has expanded enough that the last scattering surface (LSS)
has intersected itself.  The intersections appear as matching circles
embedded in opposite sides of the cosmic microwave background
radiation (CMBR).  Temperature variations along the matching circles
should be identical.  This technique has the benefit of being able to
exactly determine the cosmic topology.  As of yet, neither of these
techniques has produced a positive result.  The LSS may not have
extended enough to intersect itself.  Or, the cosmic topology may be
observable but complex and difficult to detect in current observations
of the CMBR or galactic distributions.

The global topology of the universe also has a more subtle influence
on fields.  In a spacetime with a periodic boundary condition imposed
by topology, only certain modes of a quantum field are allowed.  As a
result, the values of locally measurable quantities, such as the
vacuum energy density of the field, are shifted from the values in a
simply connected spacetime manifold.  This is called the topological
Casimir effect.  DeWitt, Hart, and Isham \cite{DeWitt} have perviously
studied the Casimir effect on the vacuum expectation value of the
stress-energy tensor, $\langle 0| T_{\mu \nu} |0\rangle$, for massless
scalar fields in several spacetimes with nontrivial topology.  In this
paper we extend their work to all possible flat manifolds, including
non-orientable manifolds.  In Section~\ref{sec:Method} we outline the
method of images to calculate $\langle 0| T_{\mu \nu} |0\rangle$ for a
free massless scalar field in a multiply-connected spacetime.  We
present the results of our calculations in
Sections~\ref{sec:OneClosed} through~\ref{sec:3-TorusKlein}.  Finally,
we summarize the effects of multiple-connectedness, orientability, and
other topological modifications on the vacuum energy density in
Section~\ref{sec:Effects}.

In this paper we assume that the universe is static, flat, and
homogeneous.  We will look at spacetime manifolds with the structure
$\mathrm{time}~(\mathbb{R}) \times \mathrm{space}~(\mathcal{M}_{3})$
where $\mathcal{M}_{3}$ is a three-dimensional spatial hypersurface
having the Euclidean space $\mathbb{R}^{3}$ as the universal covering
space.  By nontrivial topology of spacetime, we mean topology of the
space $\mathcal{M}_{3}$, not the spacetime.  Also, we will use natural
units where $G = \hbar = c = 1$, and the metric signature of $+2$.

\section{The Stress-Energy Tensor and the Method of Images}
\label{sec:Method}
Calculation of the vacuum expectation value of the stress-energy
tensor in a flat universe with nontrivial topology is relatively
straightforward.  This is due to the facts that all curvature
components vanish in a flat geometry and that topology of the manifold
appears only in an interval between two points in spacetime.  Once the
interval for a particular spacetime manifold is found, the calculation
of $\langle 0| T_{\mu \nu} |0\rangle$ reduces to simple
differentiation of the Hadamard elementary function, taking the
coincidence limit, and subtracting an appropriate infinity to
renormalize its value.

The stress-energy tensor for a massless free scalar field $\phi(x)$ is
given by
\begin{equation}
	T_{\mu \nu} = (1-2\xi)\phi_{;\mu} \phi_{;\nu} + \left( 2\xi -
	\frac{1}{2} \right) g_{\mu \nu} \phi_{;\alpha}\phi^{;\alpha} -
	2\xi \phi \phi_{; \mu \nu},
\end{equation}
where $\xi$ is the curvature coupling constant.  We will let $\xi =
\frac{1}{6}$ for conformal coupling.  The scalar field satisfies the
massless Klein-Gordon equation $\square_x\phi(x) = 0$.

Since every term in $T_{\mu \nu}$ is quadratic in the field variable
$\phi$, we can split the point $x$ into the two points $x$ and
$\tilde{x}$, and after taking covariant derivatives of $\phi$, the two
points are brought back together in the coincidence limit $\tilde{x}
\to x$:
\begin{equation}
	T_{\mu \nu} = \frac{1}{2} \lim_{\tilde{x} \to x} \left[ (1-2\xi)
	\nabla_{\mu} \widetilde{\nabla}_{\nu} + \left(2\xi -
	\frac{1}{2}\right) g_{\mu \nu} \nabla_{\alpha}
	\widetilde{\nabla}^{\alpha} - 2\xi \nabla_{\mu} \nabla_{\nu}
	\right] \{ \phi(x), \phi(\tilde{x})\}.
\end{equation}
Since there is no preference for either point, the stress-energy
tensor $T_{\mu\nu}$ is symmetrized over $x$ and $\tilde{x}$.  The
covariant derivative $\nabla_{\mu}$ and $\widetilde{\nabla}_{\nu}$ are
to be applied to $\phi(x)$ and $\phi(\tilde{x})$, respectively.  The
expectation value of the stress-energy tensor needs to be evaluated
with respect to the vacuum state in each spacetime.

The Minkowski vacuum state, denoted by $|0_{M}\rangle$, is defined in
terms of the positive frequency modes,
\begin{equation}
	\label{eq:modes}
	u_{\mathbf{k}} = \frac{e^{-i k_\alpha x^\alpha}}{\left[ 2 \omega
	(2\pi)^3\right]^{\frac{1}{2}}},
\end{equation}
which are plane-wave solutions for the Klein-Gordon equation.  The
vacuum state is the state which is annihilated by the operator
$a_{\mathbf{k}}$ for all wave vectors $\mathbf{k}$:
\begin{equation}
	a_{\mathbf{k}} |0_M \rangle = 0.
\end{equation}
The spatial part of the Minkowski space is a three-dimensional
Euclidean space, and thus there is no restriction for values of the
wave vector $\mathbf{k}$.

If a boundary condition is imposed on the space, only a set of certain
discrete modes will be allowed.  For example, if the space is periodic
in the $x$ direction, then the $x$-component of $\mathbf{k}$ will be
allowed to have only values equal to $2\pi n/L$, where $L$ is the
circumference of the space in the $x$ direction and $n$ is any
non-zero integer.  This is because a plane wave travelling around the
space in the closed direction will be out of phase with itself if its
wavelength has a value other than $L/n$, and thus will destructively
interfere with itself.  Let $\mathbf{k}_{n}$ be the allowed discrete
mode.  The vacuum state $|0\rangle$ of this multiply-connected
spacetime is defined by
\begin{equation}
	a_{\mathbf{k}_n} | 0 \rangle = 0
\end{equation}
for all allowed values of $\mathbf{k}_{n}$.  This vacuum state is
different from the Minkowski vacuum state $|0_{M}\rangle$.  Different
boundary conditions on the space manifold will result in different
vacuum states.

The vacuum expectation value of the stress-energy tensor can be
written as
\begin{equation}
	\label{eq:expectation}
	\langle 0 | T_{\mu \nu} | 0 \rangle = \frac{1}{2} \lim_{\tilde{x}
	\to x} \left[ (1-2\xi) \nabla_{\mu} \widetilde{\nabla}_{\nu} +
	\left(2\xi - \frac{1}{2}\right) g_{\mu \nu} \nabla_{\alpha}
	\widetilde{\nabla}^{\alpha} - 2\xi \nabla_{\mu} \nabla_{\nu}
	\right] D^{(1)}(x,\tilde{x}),
\end{equation}
where the Hadamard elementary function for a massless scalar field,
$D^{(1)}(x, \tilde{x})$, is defined as the vacuum expectation value of
the anticommutator of the field variables:
\begin{equation}
	D^{(1)}(x,\tilde{x}) \equiv \langle 0 | \{ \phi(x),
	\phi(\tilde{x})\} | 0 \rangle.
\end{equation}
$D^{(1)}$ satisfies the massless Klein-Gordon equation $\square_x
D^{(1)}(x,\tilde{x}) = 0$.

The vacuum stress-energy tensor as defined by
Eq.~(\ref{eq:expectation}) is infinite, and it must be renormalized by
subtracting an appropriate infinity from it.  The correct value to
subtract is given by the normal ordering of the creation and
annihilation operators, and it turns out to be the Minkowski vacuum
expectation value.  Thus, the renormalized vacuum expectation value of
the stress-energy tensor in a spacetime with nontrivial topology is
given by
\begin{equation}
	\label{eq:renormalized}
	\langle T_{\mu \nu} \rangle \equiv \langle 0 | T_{\mu \nu} | 0
	\rangle_{ren} = \langle 0 | T_{\mu \nu} | 0 \rangle - \langle 0_M
	| T_{\mu \nu} | 0_M \rangle.
\end{equation}
Notice that the normalized stress-energy tensor in Minkowski space is
identically zero.
\begin{figure*}
	\centering
	\epsfig{figure=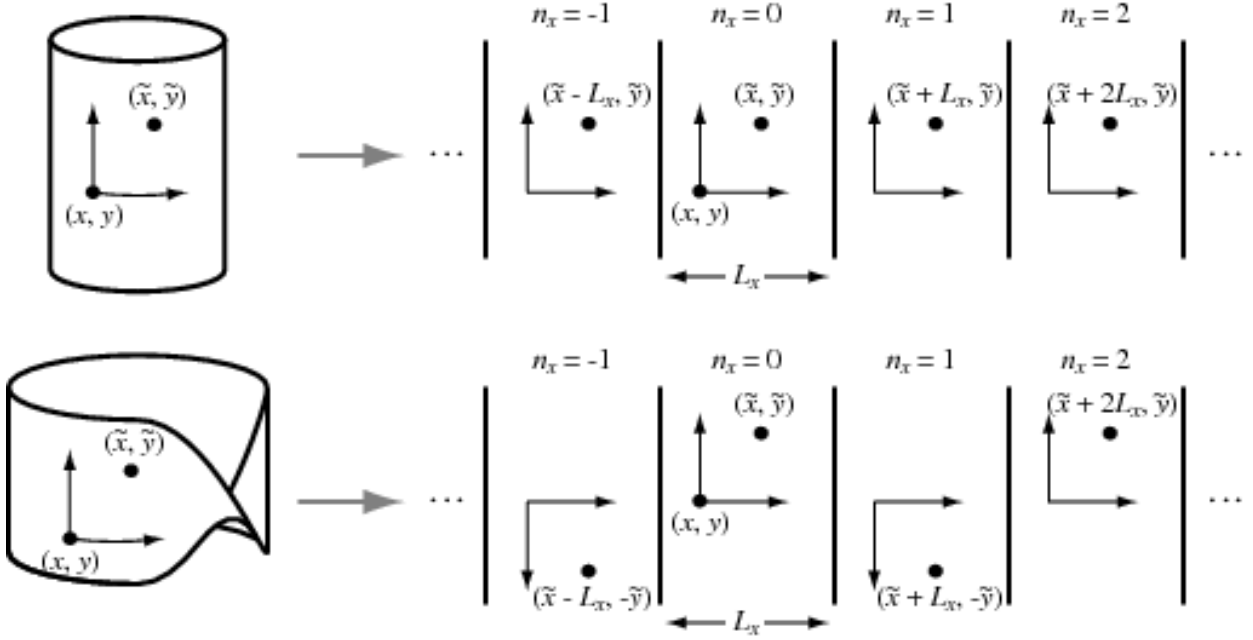, width=\textwidth}
	\caption{Unrolling a cylinder (top) and M\"{o}bius Strip (bottom)
	to determine image point locations.}
	\label{fig:UnwrappedCylinderMobius}
\end{figure*}

The Hadamard function for a massless scalar field in Minkowski space
is a simple function of the half-squared interval $\sigma$ between two
points $x$ and $\tilde{x}$:
\begin{equation}
	\label{eq:hadamard}
	D^{(1)}(x,\tilde{x}) = \frac{1}{4\pi \sigma},
\end{equation}
where $\sigma$ is defined as
\begin{equation*}
	\sigma = \frac{1}{2} [-(t - \tilde{t})^{2} + (x - \tilde{x})^{2} +
	(y - \tilde{y})^{2} + (z - \tilde{z})^{2}].
\end{equation*}

The Hadamard function for a spacetime with nontrivial topology has the
same functional form as the Minkowski case, but we replace $\sigma$ in
Eq.  (\ref{eq:hadamard}) to reflect the different periodic boundary
conditions of our chosen spacetime.  Because the spacetime is
multiply-connected, there can be more than one interval connecting two
points $(t,\ x,\ y,\ z)$ and $(\tilde{t},\ \tilde{x},\ \tilde{y},\
\tilde{z})$.  For example, suppose the spacetime is closed in the
$x$-direction.  We can connect $(t,\ x,\ y,\ z)$ and $(\tilde{t},\
\tilde{x},\ \tilde{y},\ \tilde{z})$ with a direct path, or we can
start from $(t,\ x,\ y,\ z)$ and circle around in the $x$-direction
once, twice, or an arbitrary number of times before arriving at
$(\tilde{t},\ \tilde{x},\ \tilde{y},\ \tilde{z})$.  Since each path
which wraps around cannot be deformed into another that wraps around a
different number of times, we must take into account all nonequivalent
paths by summing over them, from $n=-\infty$ to $n=\infty$, where $n$
would represent the number of loops around the space.

Equivalently, we can unwrap the space and tile copies of itself in a
way required by the topology of the space.  In each copy there is an
image of the original point $(\tilde{t},\ \tilde{x},\ \tilde{y},\
\tilde{z})$.  A path connects $(t,\ x,\ y,\ z)$ to each image of
$(\tilde{t},\ \tilde{x},\ \tilde{y},\ \tilde{z})$.  The top of
Figure~\ref{fig:UnwrappedCylinderMobius} demonstrates this technique.
The $x$-coordinates of the image points are $\tilde{x} \pm L_x,\
\tilde{x} \pm 2L_x\,...,\ \tilde{x} \pm n_x L_x$, where $L_x$ is the
circumference in the closed spatial direction.  Thus, we define
$\sigma$ for for this space as
\begin{equation}
	\label{eq:OneTorus}
	\sigma = \frac{1}{2} [-(t - \tilde{t})^{2} + (x - \tilde{x} -
	n_{x} L_{x})^{2} + (y - \tilde{y})^{2} + (z - \tilde{z})^{2}].
\end{equation}
We must sum over every contribution from each image point to construct
the Hadamard function for a particular spacetime.  

The term $n_x=0$ corresponds to the case with no boundary.  This is
the term which will give us an infinity associated with the
unrenormalized stress-energy tensor of Minkowski space and will be
subtracted in Eq.~(\ref{eq:renormalized}).  So by excluding the
$n_x=0$ term from the summation, the renormalized Hadamard function
$D^{(1)}_{ren}(x,\tilde{x})$ is obtained:
\begin{equation}
	D^{(1)}_{ren}(x,\tilde{x}) = \sideset{}{'}\sum_{n_{x}}
	D^{(1)}(\sigma) \equiv \sum_{\substack{n_{x} = -\infty \\ n_{x}
	\neq 0}}^{\infty} D^{(1)}(\sigma),
\end{equation}
where $\Sigma'$ will indicate skipping over the simply-connected case.
For intervals with multiple summation indices, this means skipping the
case where all indices are simultaneously $0$.  Finally, using
Eq.~(\ref{eq:expectation}), the renormalized stress-energy tensor
$\langle T_{\mu \nu} \rangle$ can be defined as
\begin{align}
	\label{eq:RenExpectation}
	\langle T_{\mu \nu} \rangle = \frac{1}{2} \lim_{\tilde{x} \to x}
	\bigg[ (1-2\xi) \nabla_{\mu} \widetilde{\nabla}_{\nu} + \left.
	\left(2\xi - \frac{1}{2}\right) g_{\mu \nu} \nabla_{\alpha}
	\widetilde{\nabla}^{\alpha} \right.  { } - 2\xi \nabla_{\mu}
	\nabla_{\nu} \bigg] D^{(1)}_{ren}(x,\tilde{x}).
\end{align}

The procedure for calculating $\langle T_{\mu \nu} \rangle$ can be
summarized as: (1) write an appropriate $\sigma$ (i.e. determine the
interval) for each manifold, (2) sum over the correct indices to
construct the renormalized Hadamard function, (3) apply the derivative
operator in Eq.~(\ref{eq:RenExpectation}), and (4) take the
coincidence limit as $\tilde{x} \to {x}$.  A table of the spacetimes
having the three-dimensional Euclidean space as a universal covering
space and their properties can be found in \cite{Riazuelo}.  We will
investigate each of these spaces individually.

\section{One Closed Dimension}
\label{sec:OneClosed}
We begin by modifying the Euclidean 3-Space in the simplest possible
way: closing one dimension to form a loop.  This can be constructed by
identifying a single pair of opposite faces of a rectangular box.  The
space's Fundamental Polyhedron (the most basic, simply-connected
spatial domain that we can tile to create the universal covering
space) is thus a ``slab'' of space, with a finite uniform thickness
$L_x$ in the $x$-direction, and infinite extension in the $y$- and
$z$-directions.  This space can also admit a flipping in one of the
open dimensions, resulting in a M\"{o}bius Strip.

%%%%%%%%%%%%%%%%%%%%%%%%%%%%%%%%%%%%%%%%%%%%%%%%%%%%%%%%%%%%%%%%%%%%%%%%%%%%%%%%%%%%%%%%%%%%%%%%%%%%%%%%%%%%%%
%%%%%%%%%%%%%%%%%%%%%%%%%%%%%%%%				1-Torus
%%%%%%%%%%%%%%%%%%%%%%%%%%%%%%%%%%%%%%%%%%
%%%%%%%%%%%%%%%%%%%%%%%%%%%%%%%%%%%%%%%%%%%%%%%%%%%%%%%%%%%%%%%%%%%%%%%%%%%%%%%%%%%%%%%%%%%%%%%%%%%%%%%%%%%%%%
\subsection{1-Torus ($E_{16}$ )}
The half-squared interval $\sigma$ of a simple 1-Torus is given by
Eq.~(\ref{eq:OneTorus}), where $n_x$ is an integer.  Inserting this
interval into Eq.~(\ref{eq:hadamard}) and calculating the expectation
value of the stress-energy tensor yields
\begin{equation}
	\langle T_{\mu \nu} \rangle_{E_{16}} = \frac{\pi^2}{90 {L_x}^4}
	\diag\left[-1, -3, 1, 1\right].
\end{equation}
This result was originally obtained by DeWitt and others
\cite{DeWitt}.  Note that the vacuum energy density $\rho = \langle
T_{00} \rangle_{E_{16}}$ is constant and less than zero throughout the
space.  As the circumference $L_{x}$ of the space increases, the
vacuum energy density approaches zero.  For the calculation of
$\langle T_{\mu\nu} \rangle$ for a massive scalar field with an
arbitrary curvature coupling in the 1-Torus Space, see Tanaka and
Hiscock \cite{Tanaka}.

%%%%%%%%%%%%%%%%%%%%%%%%%%%%%%%%%%%%%%%%%%%%%%%%%%%%%%%%%%%%%%%%%%%%%%%%%%%%%%%%%%%%%%%%%%%%%%%%%%%%%%%%%%%%%%
%%%%%%%%%%%%%%%%%%%%%%%%%%%%%%%%			1-Torus	with Flip
%%%%%%%%%%%%%%%%%%%%%%%%%%%%%%%%%%%%%%%%%%
%%%%%%%%%%%%%%%%%%%%%%%%%%%%%%%%%%%%%%%%%%%%%%%%%%%%%%%%%%%%%%%%%%%%%%%%%%%%%%%%%%%%%%%%%%%%%%%%%%%%%%%%%%%%%%
\subsection{1-Torus with Flip ($E_{17}$)}
If we unroll a M\"{o}bius Strip onto a flat plane, we see as in the
bottom of Figure~\ref{fig:UnwrappedCylinderMobius} an interval of
\begin{eqnarray}
	\sigma = & \frac{1}{2} \{-(t - \tilde{t})^{2} + (x - \tilde{x} -
	n_{x} L_{x})^{2} + [y - (-1)^{n_{x}} \tilde{y}]^{2} + (z -
	\tilde{z})^{2} \}^{2}.
\end{eqnarray}
Every time we slide by $L_{x}$ in the $x$ direction, the $y$
coordinate of the image point changes its sign.  This is the first of
our non-orientable manifolds.  Note that $L_x$ is defined as the
circumference of the universe, instead of the distance the field must
traverse to return to the original state.  In the latter view, the
Fundamental Polyhedron is now of width $2 L_x$.  DeWitt, Hart, and
Isham \cite{DeWitt} have shown that the vacuum stress-energy tensor in
the 1-Torus with Flip (designated as $E_{17}$ in \cite{Riazuelo}) is
\begin{align}
	\langle T_{\mu \nu} \rangle_{E_{17}} = \frac{1}{16} \langle T_{\mu
	\nu} \rangle_{E_{16}} + \langle T_{\mu \nu} \rangle_{flip} (y),
\end{align}
where
\begin{align*}
	\langle T_{\mu \nu} \rangle_{flip} (y) = \frac{2}{3 \pi^2}
	\sum_{\substack{n_x\\n_x\ odd}} \frac{{L_x}^2 {n_x}^2}{({L_x}^2
	{n_x}^2 + 4y^2 )^3} \cdot \diag\left[-1, -2, 0, 1 \right],
\end{align*}
which can be written in terms of elementary functions:
\begin{align*}
	\langle T_{\mu \nu} \rangle_{flip} (y) = & \frac{1}{192 \pi
	{L_x}^3 y^3} \left\{ {L_x}^2 \tanh \left( \frac{\pi y}{L_x}
	\right) - \pi y \sech^2 \left( \frac{\pi y}{L_x} \right) \left[
	L_x - 2 \pi y \tanh \left( \frac{\pi y}{L_x} \right) \right]
	\right\} \\
	& \quad \times \diag\left[-1, -2, 0, 1 \right].
\end{align*}

\begin{figure}
	\centering
	\epsfig{file=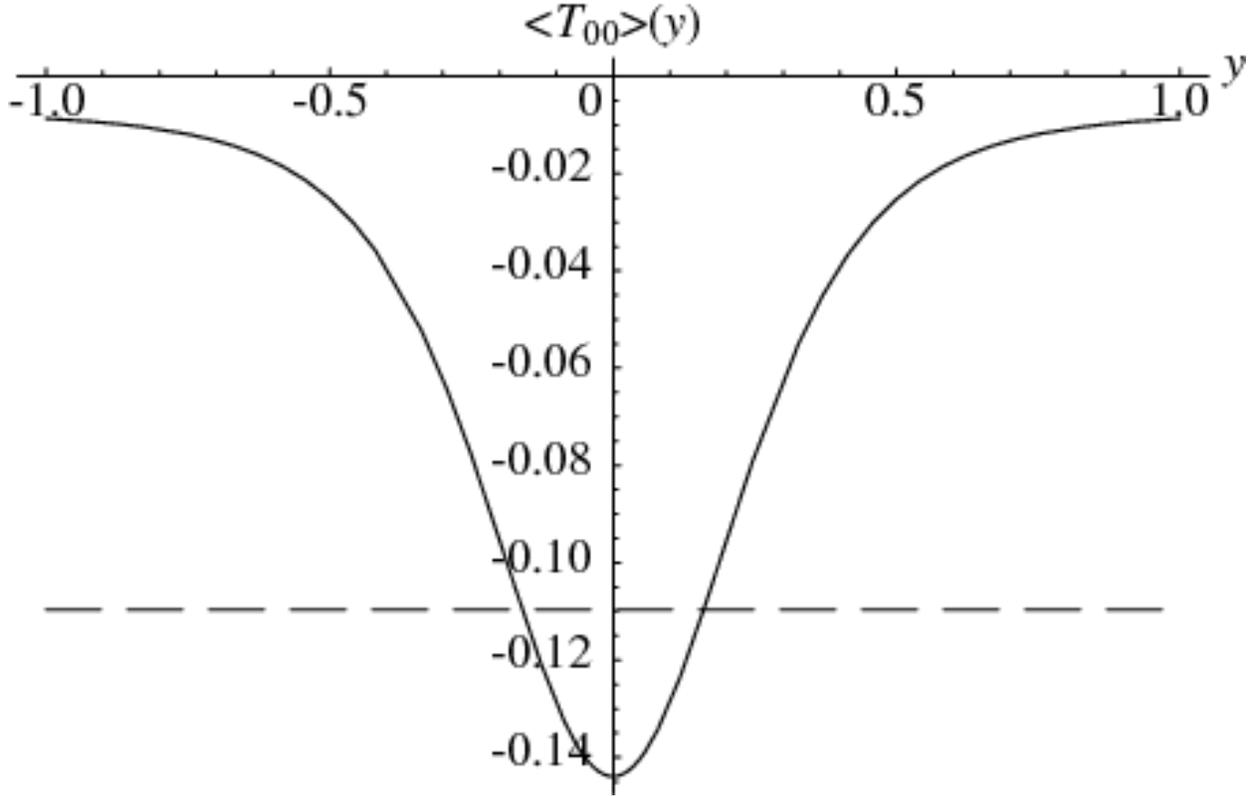,width=\textwidth} 	
	\caption{The solid line is the vacuum energy density in 1-Torus
	with Flip.  The dashed line indicates the constant energy density
	value for the 1-Torus.  The $y$-axis extends to infinity in either
	direction.}
	\label{plot:E17}
\end{figure}

Because the field must traverse the universe twice in the
$x$-direction to have the same value, the circumference of the
universe is effectively doubled, and the constant shift in the
stress-energy is only $1/16$ of that for the 1-Torus Space without
flip.  Also, $\langle T_{\mu\nu} \rangle$ is now $y$-dependent (see
Fig.~\ref{plot:E17}).

\section{Two Closed Dimensions}
Now, we close the space in two directions, $x$ and $y$, but let it
extend to infinity in the $z$-direction.  The space is a stack of
rectangular columns of depth $L_{x}$, width $L_{y}$, and infinite
height in the $z$-direction.  While stacking, four transformations can
be introduced: horizontal flip, vertical flip, half-turn, and
half-turn with flip.  In these spaces, each $z = const$ slice will be
either a 2-Torus or a Klein bottle.

%%%%%%%%%%%%%%%%%%%%%%%%%%%%%%%%%%%%%%%%%%%%%%%%%%%%%%%%%%%%%%%%%%%%%%%%%%%%%%%%%%%%%%%%%%%%%%%%%%%%%%%%%%%%%%
%%%%%%%%%%%%%%%%%%%%%%%%%%%%%%%%				2-Torus
%%%%%%%%%%%%%%%%%%%%%%%%%%%%%%%%%%%%%%%%%%
%%%%%%%%%%%%%%%%%%%%%%%%%%%%%%%%%%%%%%%%%%%%%%%%%%%%%%%%%%%%%%%%%%%%%%%%%%%%%%%%%%%%%%%%%%%%%%%%%%%%%%%%%%%%%%
\subsection{2-Torus ($E_{11}$)}
The half-squared interval of the simple 2-Torus is given by
\begin{eqnarray}
	\sigma = \frac{1}{2} [-(t - \tilde{t})^{2} + (x - \tilde{x} -
	n_{x} L_{x})^{2} + (y - \tilde{y} - n_{y} L_{y})^{2} + (z -
	\tilde{z})^{2} ]^{2}.
\end{eqnarray}
The expectation value for this space is
\begin{equation}
	\langle T_{\mu \nu} \rangle_{E_{11}} =
	\sideset{}{'}\sum_{\substack{n_x,\ n_y}} [\text{2-Torus}],
\end{equation}
where
\begin{align}
	[\text{2-Torus}] = & \frac{1}{2 \pi^2} \frac{1}{({L_x}^2 {n_x}^2 +
	{L_y}^2 {n_y}^2)^3} \nonumber \\
	& \quad \times \left( {L_x}^2 {n_x}^2 \diag\left[-1, -3, 1,
	1\right] + {L_y}^2 {n_y}^2 \diag\left[-1, 1, -3, 1\right] \right).
\end{align}
Although this summation is not expressible in terms of elementary
functions, it is convergent and results in a uniform negative shift in
the vacuum energy level throughout the space.

%%%%%%%%%%%%%%%%%%%%%%%%%%%%%%%%%%%%%%%%%%%%%%%%%%%%%%%%%%%%%%%%%%%%%%%%%%%%%%%%%%%%%%%%%%%%%%%%%%%%%%%%%%%%%%
%%%%%%%%%%%%%%%%%%%%%%%%%%%%%%%%	2-Torus	with Vertical Flip
%%%%%%%%%%%%%%%%%%%%%%%%%%%%%%%%%%%%%%%%%%
%%%%%%%%%%%%%%%%%%%%%%%%%%%%%%%%%%%%%%%%%%%%%%%%%%%%%%%%%%%%%%%%%%%%%%%%%%%%%%%%%%%%%%%%%%%%%%%%%%%%%%%%%%%%%%
\subsection{2-Torus with Vertical Flip ($E_{13}$)}
For this space, the vertical (non-closed) dimension, $z$, is flipped
when gluing opposite faces in the $x$-direction.  The interval for
this space is
\begin{equation}
	\sigma = \frac{1}{2} \{ -(t - \tilde{t})^{2} + (x - \tilde{x} -
	n_{x} L_{x})^{2} + (y - \tilde{y} - n_{y} L_{y})^{2} + [z -
	(-1)^{n_{x}} \tilde{z}]^{2} \},
\end{equation}
and the resulting stress-energy tensor has the form
\begin{align}
	\langle T_{\mu \nu} \rangle_{E_{13}} = \sideset{}{'}\sum_{n_x\
	even,\ n_{y}} [\text{2-Torus}] + \sum_{n_x\ odd,\ n_{y}}
	[\text{V-Flip}(z)],
\end{align}
where
\begin{align*}
	[\text{V-Flip}(z)] = & \frac{2}{3 \pi^2} \frac{1}{({L_x}^2 {n_x}^2
	+ {L_y}^2 {n_y}^2 + 4z^2)^3} \\
	& \quad \times \left( {L_x}^2 {n_x}^2 \diag\left[-1, -2, 1, 0
	\right] + {L_y}^2 {n_y}^2 \diag\left[-1, 1, -2, 0 \right] \right).
\end{align*}

\begin{figure*}
	\centering
	\mbox{\subfigure[2-Torus with Vertical
	Flip]{\scalebox{0.6}{\epsfig{figure=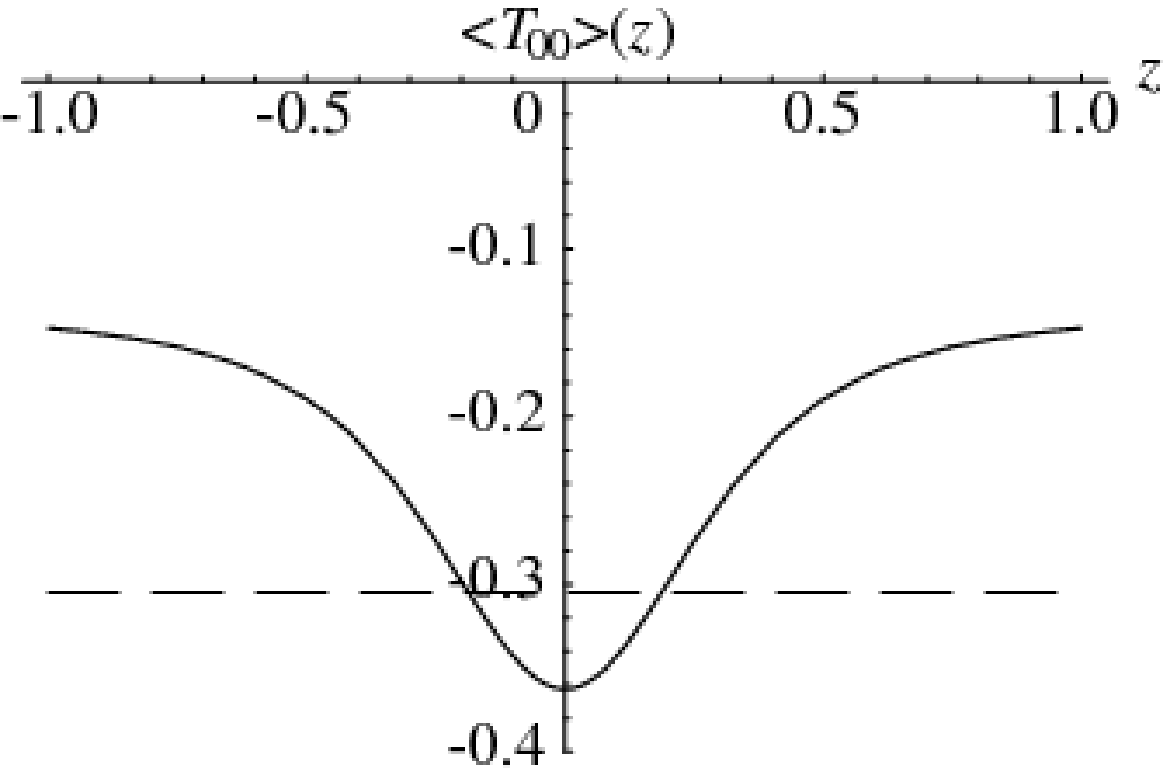}}} \quad
	\subfigure[2-Torus with Horizontal
	Flip]{\scalebox{0.6}{\epsfig{figure=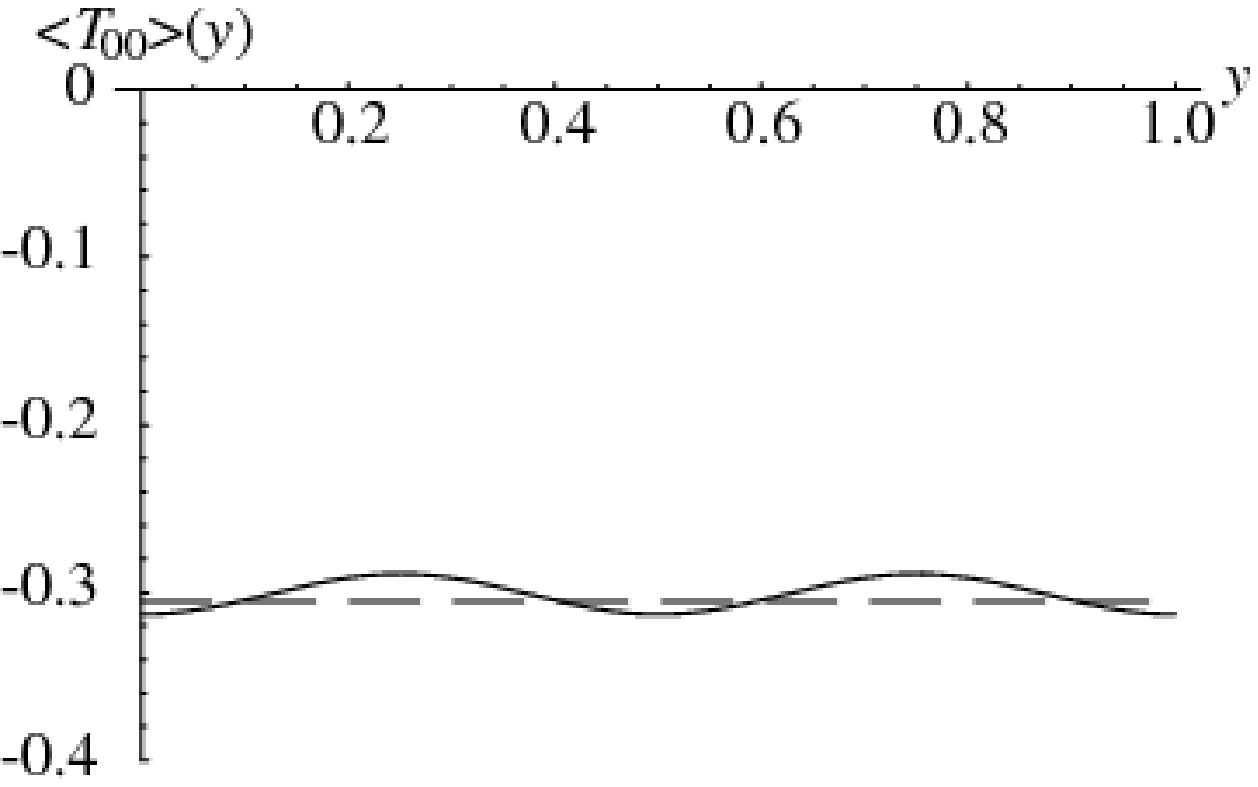}}} }
	\caption{Vacuum energy densities in the 2-Torus with Vertical and
	Horizontal Flips, with $L_x = L_y = L_z = 1$.  The dashed line is
	the value for the plain 2-Torus.  The regions are $x=const$ slices
	of the Fundamental Polyhedra.  The $z$-axis extends to infinity in
	either direction.}
	\label{plot:CompareE11E13E14}
\end{figure*}

The first summation corresponds to the stress-energy for the 2-Torus.
This summation is exactly the same as the ordinary 2-Torus case if the
interval in the $x$-direction is doubled.  The $z$-dependence appears
in the summation over odd values of $n_{x}$.  Since $z$ is not a
closed dimension, the vacuum energy density is a single well that is
most negative in the $z = 0$ plane.  The plot of the vacuum energy
density is shown in Figure~\ref{plot:CompareE11E13E14} (a).

%%%%%%%%%%%%%%%%%%%%%%%%%%%%%%%%%%%%%%%%%%%%%%%%%%%%%%%%%%%%%%%%%%%%%%%%%%%%%%%%%%%%%%%%%%%%%%%%%%%%%%%%%%%%%%
%%%%%%%%%%%%%%%%%%%%%%%%%%%%%%%%	2-Torus	with Horizontal Flip
%%%%%%%%%%%%%%%%%%%%%%%%%%%%%%%%%%%%%%%%%%
%%%%%%%%%%%%%%%%%%%%%%%%%%%%%%%%%%%%%%%%%%%%%%%%%%%%%%%%%%%%%%%%%%%%%%%%%%%%%%%%%%%%%%%%%%%%%%%%%%%%%%%%%%%%%%
\subsection{2-Torus with Horizontal Flip ($E_{14}$)}
As we translate in $x$ we may instead choose to flip in the
horizontal, or closed dimension, $y$.  Then the interval becomes
\begin{equation}
	\sigma = \frac{1}{2} \{ -(t - \tilde{t})^{2} + (x - \tilde{x} -
	n_{x} L_{x})^{2} + [y - (-1)^{n_{x}} \tilde{y} - n_{y} L_{y} ]^{2}
	+ (z - \tilde{z})^{2} \}.
\end{equation}

The vacuum stress-energy tensor is now
\begin{equation}
	\langle T_{\mu \nu} \rangle_{E_{14}} = \sideset{}{'}\sum_{n_x\
	even,\ n_{y}} [\text{2-Torus}] + \sum_{n_x\ odd,\ n_{y}}
	[\text{H-Flip}(y)],
\end{equation}
with
\begin{equation*}
	[\text{H-Flip}(y)] = \frac{2}{3 \pi^2} \frac{{L_x}^2
	{n_x}^2}{[{L_x}^2 {n_x}^2 + (2y - L_y n_y)^2 ]^3} \diag\left[-1,
	-2, 0, 1 \right].
\end{equation*}
Since $y$ is a closed dimension, a periodic energy density well
appears in that direction (see Figure~\ref{plot:CompareE11E13E14}
(b)).

%%%%%%%%%%%%%%%%%%%%%%%%%%%%%%%%%%%%%%%%%%%%%%%%%%%%%%%%%%%%%%%%%%%%%%%%%%%%%%%%%%%%%%%%%%%%%%%%%%%%%%%%%%%%%%
%%%%%%%%%%%%%%%%%%%%%%%%%%%%%%%%	2-Torus	with Half-Turn
%%%%%%%%%%%%%%%%%%%%%%%%%%%%%%%%%%%%%%%%%%
%%%%%%%%%%%%%%%%%%%%%%%%%%%%%%%%%%%%%%%%%%%%%%%%%%%%%%%%%%%%%%%%%%%%%%%%%%%%%%%%%%%%%%%%%%%%%%%%%%%%%%%%%%%%%%
\subsection{2-Torus with Half-Turn ($E_{12}$)}
The vertical and horizontal flip can be combined to generate a
half-turn corkscrew motion when opposite faces in the $x$-direction
are glued.  Aside from the plain 2-Torus, this is the only orientable
manifold with 2 closed dimensions.  Combining the intervals of the two
previous flips gives
\begin{equation}
	\sigma = \frac{1}{2} \{ -(t - \tilde{t})^{2} + (x - \tilde{x} -
	n_{x} L_{x})^{2} + [y - (-1)^{n_{x}} \tilde{y} - n_{y} L_{y}]^{2}
	+ [z - (-1)^{n_{x}} \tilde{z} ]^{2} \}.
\end{equation}

\begin{figure}
	\epsfig{file=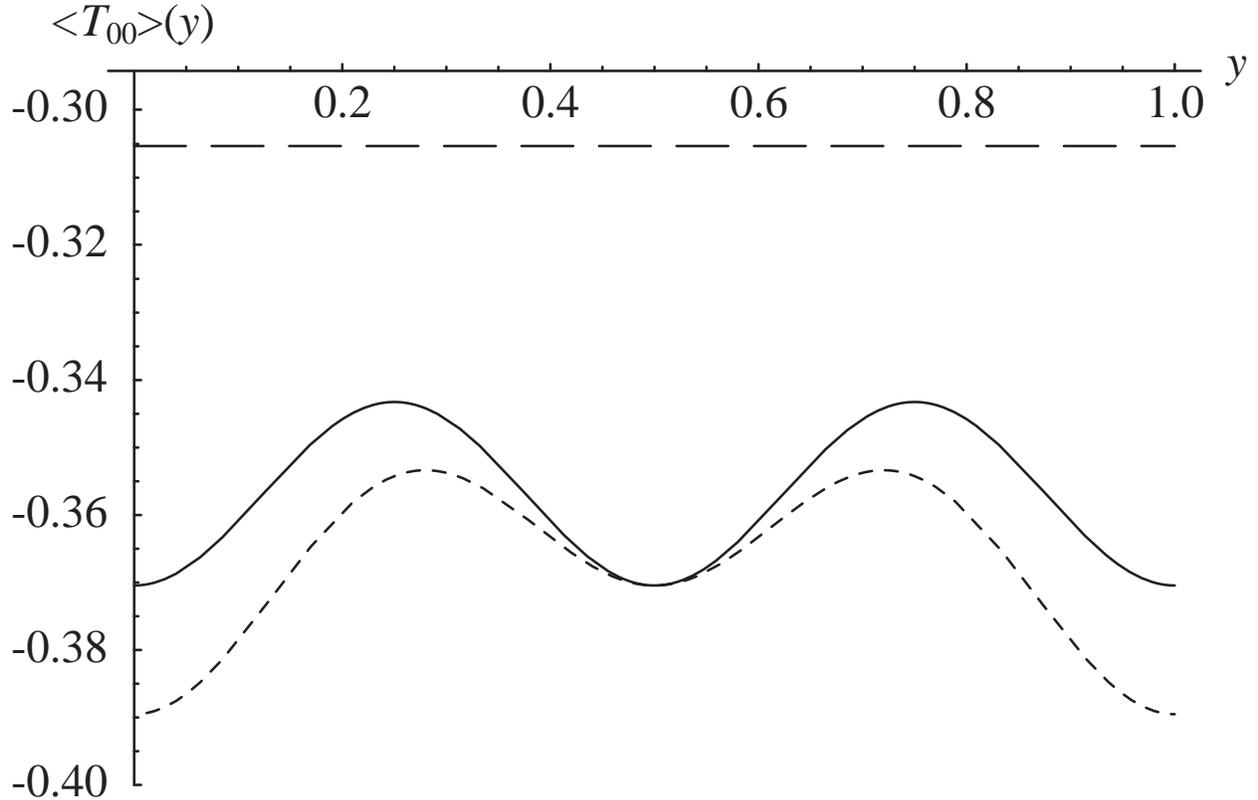,width=\textwidth}
	\caption{Adding a flip: comparison of 2-Torus with Half-Turn
	(solid) and 2-Torus with Half-Turn and Flip (small dashed).
	Additionally, energy density of the plain 2-Torus (long dashed) is
	shown.  The region is a single $x=const$ and $z=0$ slice of the
	Fundamental Polyhedron.}
	\label{plot:compareE12E15}
\end{figure}

The resulting expectation value is also a combination of the two
flips:
\begin{equation}
	\langle T_{\mu \nu} \rangle_{E_{12}} = \sideset{}{'}\sum_{n_x\
	even,\ n_{y}} [\text{2-Torus}] + \sum_{n_x\ odd,\ n_{y}}
	[\text{Turn}(y,z)],
\end{equation}
where
\begin{align*}
	[\text{Turn} (y,z)] = & \frac{1}{6 \pi^2} \frac{1}{[{L_x}^2
	{n_x}^2 + (2y - L_y n_y)^2 + 4z^2]^3} \\
	& \quad \times \Big\{ {L_x}^2 {n_x}^2 \diag\left[-5, -7, 1, 1
	\right] + (2y - L_{y} n_{y})^{2} \diag\left[-1, 1, 1, -3 \right]
	\\
	& \quad + 4z^2 \diag\left[-1, 1, -3, 1 \right] - 8 (2y - L_{y}
	n_{y}) z \left(\delta^{y}_{\ \mu} \delta^{z}_{\ \nu} +
	\delta^{z}_{\ \mu} \delta^{y}_{\ \nu}\right) \Big\}.
\end{align*}
There are now non-zero off-diagonal terms in $\langle T_{\mu\nu}
\rangle$.  In fact, we find off-diagonal elements every time a space
is rotated.  As expected, we have periodic behavior in $y$ and a
single well in $z$.  The energy density in this space is shown in
Figure~\ref{plot:compareE12E15}.

%%%%%%%%%%%%%%%%%%%%%%%%%%%%%%%%%%%%%%%%%%%%%%%%%%%%%%%%%%%%%%%%%%%%%%%%%%%%%%%%%%%%%%%%%%%%%%%%%%%%%%%%%%%%%%
%%%%%%%%%%%%%%%%%%%%%%%%%%%%	2-Torus	with Half-Turn and Flip
%%%%%%%%%%%%%%%%%%%%%%%%%%%%%%%%%%%%%%%%%%
%%%%%%%%%%%%%%%%%%%%%%%%%%%%%%%%%%%%%%%%%%%%%%%%%%%%%%%%%%%%%%%%%%%%%%%%%%%%%%%%%%%%%%%%%%%%%%%%%%%%%%%%%%%%%%
\subsection{2-Torus with Half-Turn and Flip ($E_{15}$)}
We now combine all corkscrew and flipping motions to form a
non-orientable manifold with interval
\begin{equation}
	\sigma = \frac{1}{2} \{ -(t - \tilde{t})^{2} + (x - \tilde{x} -
	n_{x} L_{x})^{2} + [y - (-1)^{n_{x}} \tilde{y} - n_{y} L_{y} ]^{2}
	+ [z - (-1)^{n_{x}} (-1)^{n_{y}} \tilde{z} ]^{2} \}.
\end{equation}
The resulting stress-energy tensor can be split into four separate
summations corresponding to terms appearing in the stress-energy in
simpler spaces.  For example, we notice that if both $n_x$ and $n_y$
are even, we recover the interval for the 2-Torus.  If both indices
are odd, we get the interval for the 2-Torus with Half-Turn.  So we
can build the stress-energy tensor as:
\begin{align}
	\langle T_{\mu \nu} \rangle_{E_{14}} = & \sideset{}{'}\sum_{n_x\
	even,\ n_y\ even} [\text{2-Torus}] + \sum_{n_x\ odd,\ n_y\ even}
	[\text{Turn}(y,z)] \nonumber \\
	& \quad {}+ \sum_{n_x\ even,\ n_y\ odd} [\text{V-Flip}(z)] +
	\sum_{n_x\ odd,\ n_y\ odd} [\text{H-Flip}(y)].
\end{align}
The energy density in this space is shown in
Figure~\ref{plot:compareE12E15}.

\section{Three Closed Dimensions -- Toroidal Cross-Section}

%%%%%%%%%%%%%%%%%%%%%%%%%%%%%%%%%%%%%%%%%%%%%%%%%%%%%%%%%%%%%%%%%%%%%%%%%%%%%%%%%%%%%%%%%%%%%%%%%%%%%%%%%%%%%%
%%%%%%%%%%%%%%%%%%%%%%%%%%%%%%%%%%%%			3-Torus
%%%%%%%%%%%%%%%%%%%%%%%%%%%%%%%%%%%%%%%%%%
%%%%%%%%%%%%%%%%%%%%%%%%%%%%%%%%%%%%%%%%%%%%%%%%%%%%%%%%%%%%%%%%%%%%%%%%%%%%%%%%%%%%%%%%%%%%%%%%%%%%%%%%%%%%%%
\subsection{3-Torus ($E_{1}$)}
The Fundamental Polyhedron is a rectangular box of width $L_y$, depth
$L_x$, and height $L_z$.  All pairs of opposite faces are identified.
The interval is
\begin{equation}
	\sigma = \frac{1}{2} [-(t - \tilde{t})^{2} + (x - \tilde{x} -
	n_{x} L_{x})^{2} + (y - \tilde{y} - n_{y} L_{y})^{2} + (z -
	\tilde{z} - n_{z} L_{z})^{2}].
\end{equation}

DeWitt, Hart, and Isham have shown that the resulting stress-energy is
again constant, and it is repeated here:
\begin{equation}
	\langle T_{\mu \nu} \rangle_{E_{1}} =
	\sideset{}{'}\sum_{\substack{n_x,\ n_y,\ n_z}} [\text{3-Torus}],
\end{equation}
where
\begin{align*}
	[\text{3-Torus}] = & \frac{1}{2 \pi^2} \frac{1}{({L_x}^2 {n_x}^2 +
	{L_y}^2 {n_y}^2 + {L_z}^2 {n_z}^2)^3} \nonumber \\
	& \quad \times \left( {L_x}^2 {n_x}^2 \diag\left[-1, -3, 1,
	1\right] + {L_y}^2 {n_y}^2 \diag\left[-1, 1, -3, 1\right] \right.
	\nonumber \\
	& \quad + \left.  {L_z}^2 {n_z}^2 \diag\left[-1, 1, 1, -3\right]
	\right).
\end{align*}
The energy density is more negative than its 2- and 1-Torus analogs.
The corresponding calculation for a massive scalar field in the
3-Torus Space can be found in \cite{Tanaka}.

%%%%%%%%%%%%%%%%%%%%%%%%%%%%%%%%%%%%%%%%%%%%%%%%%%%%%%%%%%%%%%%%%%%%%%%%%%%%%%%%%%%%%%%%%%%%%%%%%%%%%%%%%%%%%%
%%%%%%%%%%%%%%%%%%%%%%%%%%%%%%%%%%%%		Half-Turn Space
%%%%%%%%%%%%%%%%%%%%%%%%%%%%%%%%%%%%%%%%%%
%%%%%%%%%%%%%%%%%%%%%%%%%%%%%%%%%%%%%%%%%%%%%%%%%%%%%%%%%%%%%%%%%%%%%%%%%%%%%%%%%%%%%%%%%%%%%%%%%%%%%%%%%%%%%%
\subsection{Half-Turn Space ($E_{2}$)}
When stacking boxes in the $x$-direction, we may flip in both $y$- and
$z$-directions resulting in a rotation by $180^{\circ}$ about the
$x$-axis.  The depth of the Fundamental Polyhedron is now $2L_x$, and
the height and width remain the same.  The interval in this orientable
manifold is
\begin{align}
	\sigma = & \frac{1}{2} \{ -(t - \tilde{t})^{2} + (x - \tilde{x} -
	n_{x} L_{x})^{2} + [y - (-1)^{n_{x}} \tilde{y} - n_{y} L_{y}]^{2}
	\nonumber \\
	& \quad {}+ [z - (-1)^{n_{x}} \tilde{z} - n_{z} L_{z}]^{2} \}.
\end{align} 

\begin{figure*}
	\centering
	\mbox{\subfigure[Half-Turn 
	Space]{\scalebox{0.45}{\epsfig{figure=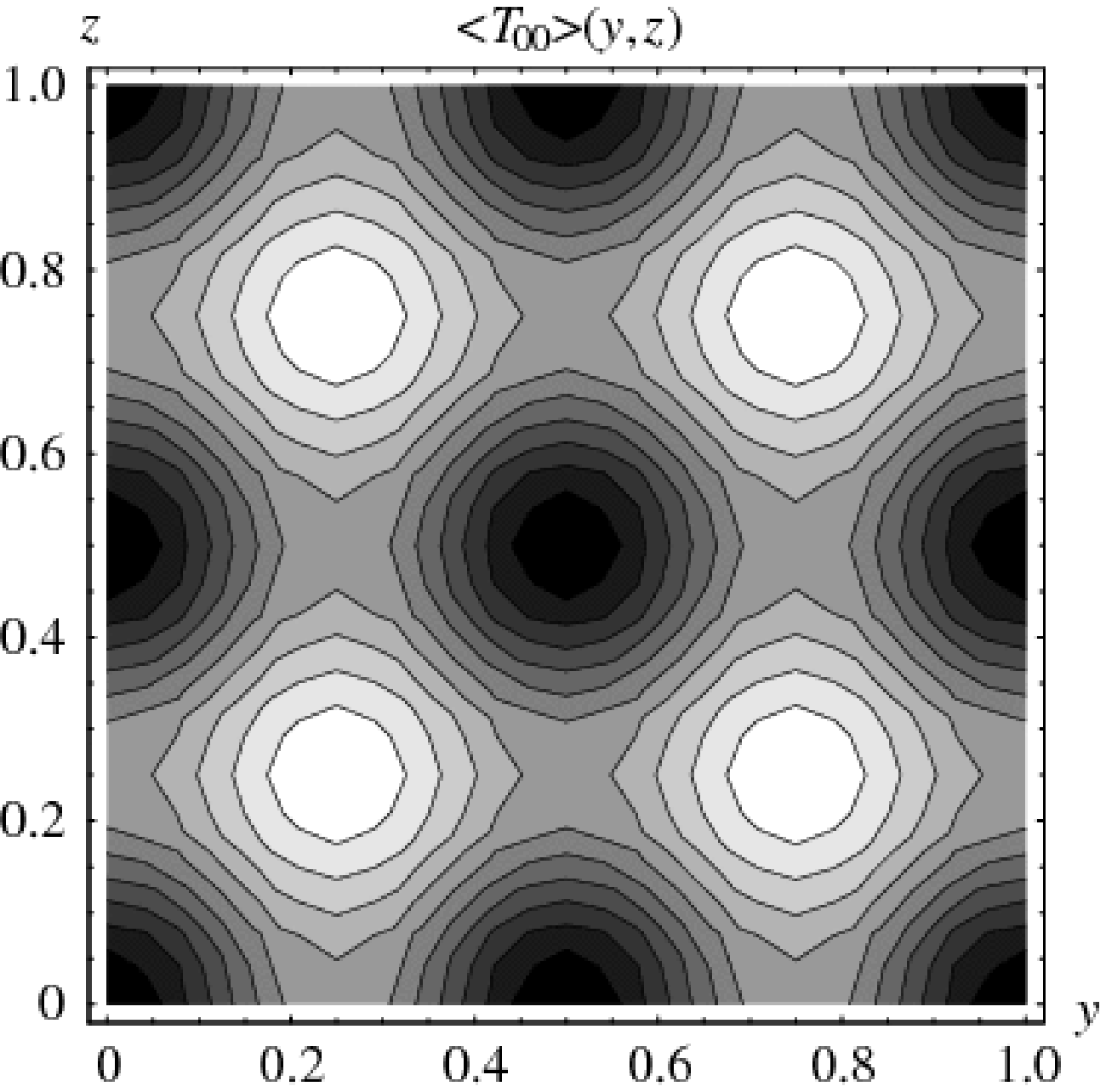}}} \quad
	\subfigure[Quarter-Turn 
	Space]{\scalebox{0.45}{\epsfig{figure=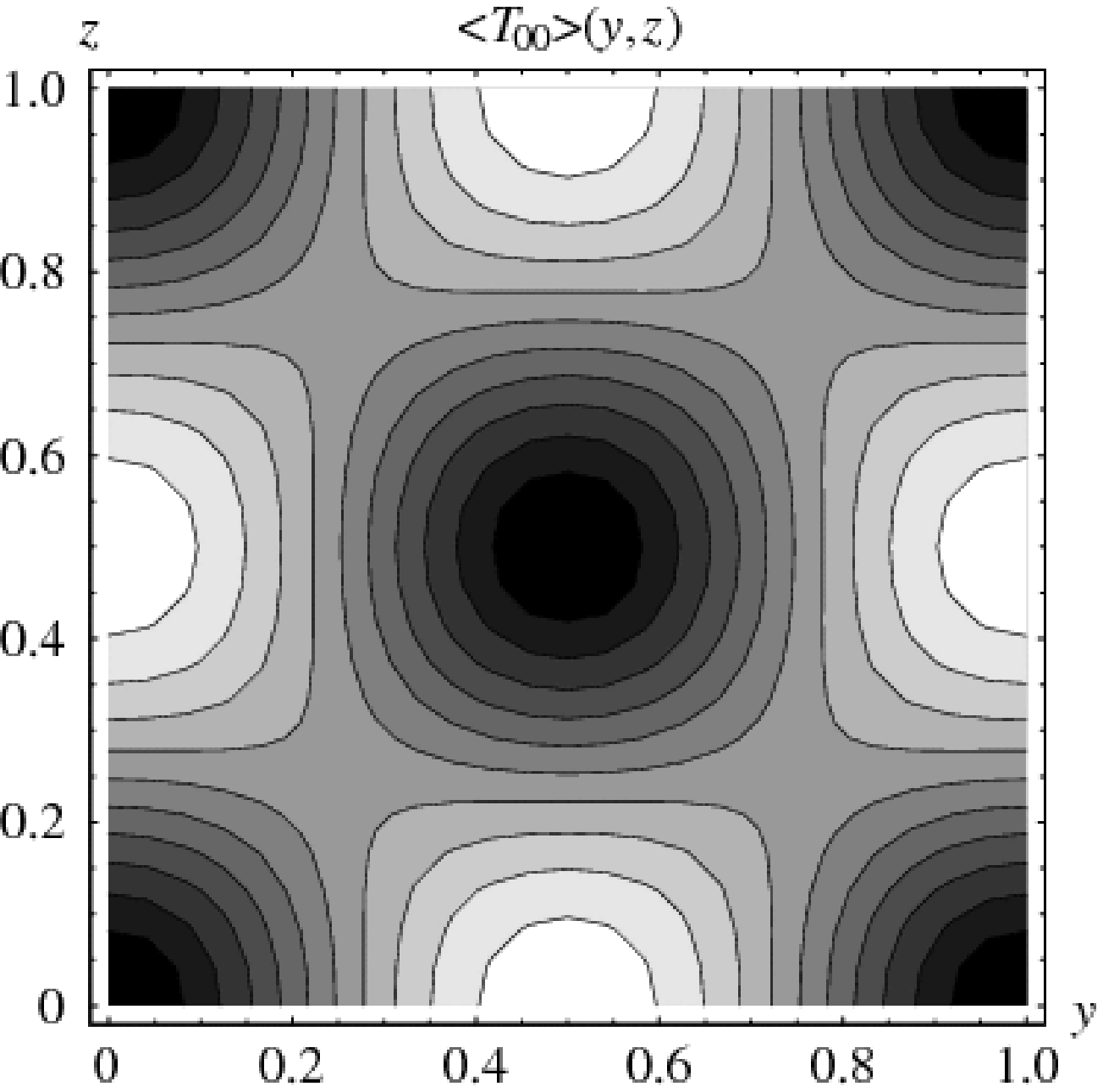}}} }	
	\caption{Vacuum energy densities in the Half-Turn Space and the
	Quarter-Turn Space with $L_x = L_y = L_z = L = 1$.  The regions
	are $x=const$ slices of the Fundamental Polyhedron.  Darker
	shading indicates more negative values.}
	\label{plot:E2E3}
\end{figure*}

Proceeding by splitting into even and odd summations gives
\begin{align}
	\langle T_{\mu \nu} \rangle_{E_{2}} =
	\sideset{}{'}\sum_{\substack{n_x\ even \\ n_{y},\ n_{z}}}
	[\text{3-Torus}] + \sum_{\substack{n_x\ odd \\ n_{y},\ n_{z}}}
	[\text{1/2-Turn}(y,z)],
\end{align}
with
\begin{align*}
	[\text{1/2-Turn} (y,z)] = & \frac{1}{6 \pi^2} \frac{1}{[{L_x}^2
	{n_x}^2 + (2y - L_y n_y)^2 + (2z - L_z n_z)^2]^3} \\
	& \quad \times \Big\{ {L_x}^2 {n_x}^2 \diag\left[-5, -7, 1, 1
	\right] + (2y - L_{y} n_{y})^{2} \diag\left[-1, 1, 1, -3 \right]
	\\
	& \quad {} + (2z - L_{z} n_{z})^{2} \diag\left[-1, 1, -3, 1
	\right] \\
	& \quad {}- 4 (2y - L_{y} n_{y}) (2z - L_{z} n_{z})
	\left(\delta^{y}_{\ \mu} \delta^{z}_{\ \nu} + \delta^{z}_{\ \mu}
	\delta^{y}_{\ \nu}\right) \Big\}.
\end{align*}
Once again we see non-zero off-diagonal elements due to the corkscrew
motion of this space.  Since flipping occurs in both the $y$- and
$z$-directions, we see in Figure~\ref{plot:E2E3} (a) periodic behavior
in the vacuum energy density in both directions.

%%%%%%%%%%%%%%%%%%%%%%%%%%%%%%%%%%%%%%%%%%%%%%%%%%%%%%%%%%%%%%%%%%%%%%%%%%%%%%%%%%%%%%%%%%%%%%%%%%%%%%%%%%%%%%
%%%%%%%%%%%%%%%%%%%%%%%%%%%%%%%%%%%%		Quarter-Turn Space
%%%%%%%%%%%%%%%%%%%%%%%%%%%%%%%%%%%%%%%%%%
%%%%%%%%%%%%%%%%%%%%%%%%%%%%%%%%%%%%%%%%%%%%%%%%%%%%%%%%%%%%%%%%%%%%%%%%%%%%%%%%%%%%%%%%%%%%%%%%%%%%%%%%%%%%%%
\subsection{Quarter-Turn Space ($E_{3}$)}
If the width and height of the rectangular box are equal ($L_{y} =
L_{z} = L$), a quarter-turn, or $90^{\circ}$, corkscrew motion is
allowed.  This is another orientable manifold.  The interval is now
\begin{align}
	\sigma = & \frac{1}{2} \bigg\{ - (t - \tilde{t})^{2} + (x -
	\tilde{x} - n_{x} L_{x})^{2} \nonumber \\
	& \quad {} + \left[ y - \cos \left( \frac{n_{x} \pi}{2} \right)
	\tilde{y} + \sin \left( \frac{n_{x} \pi}{2} \right) \tilde{z} -
	n_{y} L \right]^{2} \nonumber \\
	& \quad {} + \left[ z - \sin \left( \frac{n_{x} \pi}{2}
	\right) \tilde{y} - \cos \left( \frac{n_{x} \pi}{2} \right)
	\tilde{z} - n_{z} L \right]^{2} \bigg\}.
\end{align}
An explicit analytical expression for the stress-energy tensor in this
space is omitted in favor of a numerical plot (see
Figure~\ref{plot:E2E3} (b)).  Since it takes four complete traversals
in the $x$-direction to return to the original axes orientation, we
find the frequency of periodicity of the energy density to be half
that of the Half-Turn Space.

%%%%%%%%%%%%%%%%%%%%%%%%%%%%%%%%%%%%%%%%%%%%%%%%%%%%%%%%%%%%%%%%%%%%%%%%%%%%%%%%%%%%%%%%%%%%%%%%%%%%%%%%%%%%%%
%%%%%%%%%%%%%%%%%%%%%%%%%%%%%%%%%%%%		Third-Turn Space
%%%%%%%%%%%%%%%%%%%%%%%%%%%%%%%%%%%%%%%%%%
%%%%%%%%%%%%%%%%%%%%%%%%%%%%%%%%%%%%%%%%%%%%%%%%%%%%%%%%%%%%%%%%%%%%%%%%%%%%%%%%%%%%%%%%%%%%%%%%%%%%%%%%%%%%%%
\subsection{Third-Turn Space ($E_{4}$)}
\begin{figure*}
	\centering
	\mbox{\subfigure[Third-Turn 
	Space]{\scalebox{0.42}{\epsfig{figure=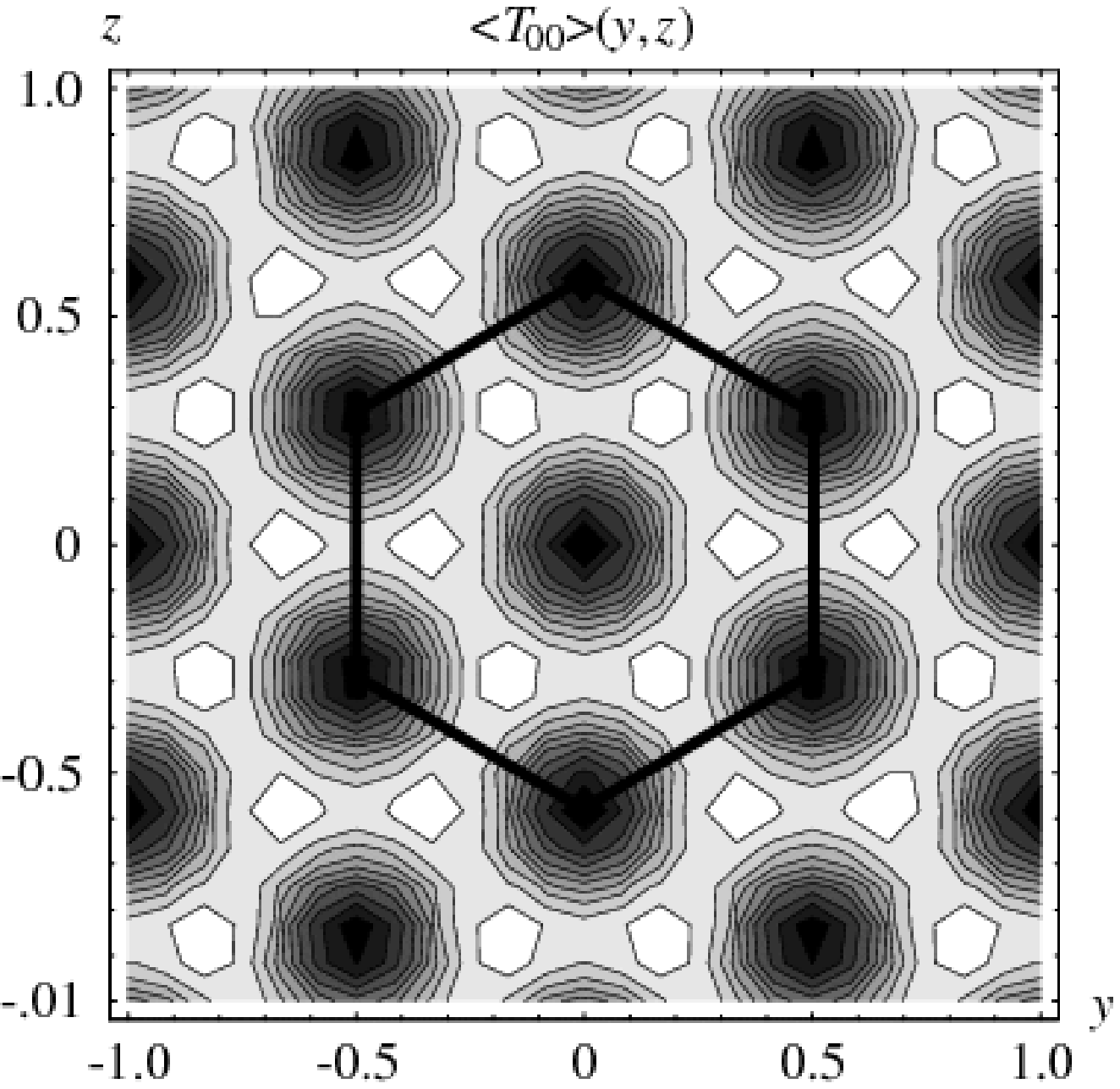}}} \quad
	\subfigure[Sixth-Turn 
	Space]{\scalebox{0.42}{\epsfig{figure=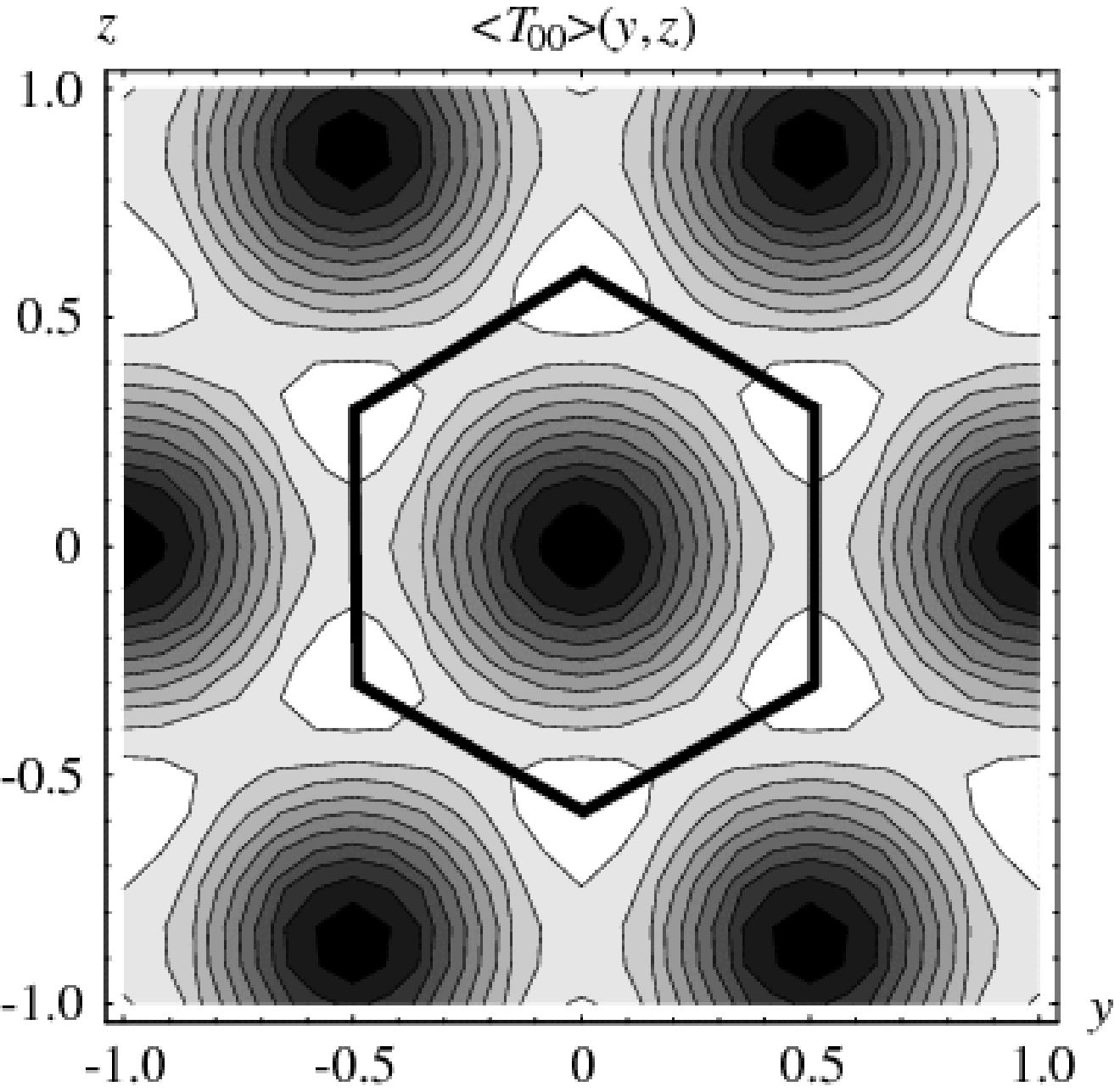}}} }	
	\caption{Vacuum energy densities in the Third-Turn Space and the
	Sith-Turn Space with $L_x = L = 1$.  The heavy black line
	indicates the region of a $x=const$ slice of the Fundamental
	Polyhedron.}
	\label{plot:E4E5}
\end{figure*}

It is possible for the $y$- and $z$-axes to rotate by $120^{\circ}$ as
we translate in the $x$-direction by $L_{x}$.  In such a case, the
Fundamental Polyhedron is a hexagonal prism.  Within this space, each
$x = const$ slice will be a hexagonal torus.  Let the distance
between base points (the distance from one rectangular face to the
opposite side) be $L$, and the height of each prism layer to be $L_x$.
The depth of this Fundamental Polyhedron is thus $3L_x$.  The
spacetime remains orientable.  The interval is now
\begin{align}
	\sigma = & \frac{1}{2} \Bigg\{ - (t - \tilde{t})^{2} + (x -
	\tilde{x} - n_{x} L_{x} )^{2} \nonumber \\
	& \quad {}+ \left[ y - \cos \left( \frac{2n_{x} \pi}{3} \right)
	\tilde{y} + \sin \left( \frac{2n_{x} \pi}{3} \right) \tilde{z} -
	n_{y} L - \frac{1}{2} n_{z} L \right]^{2} \nonumber \\
	& \quad {}+ \left[ z - \sin \left( \frac{2n_{x} \pi}{3}
	\right) \tilde{y} - \cos \left( \frac{2n_{x} \pi}{3} \right)
	\tilde{z} - \frac{\sqrt{3}}{2} n_{z} L \right]^{2} \Bigg\}.
\end{align}
The numerical result is shown in Figure~\ref{plot:E4E5} (a).  The
energy density exhibits the hexagonal structure of the underlying
topology.

%%%%%%%%%%%%%%%%%%%%%%%%%%%%%%%%%%%%%%%%%%%%%%%%%%%%%%%%%%%%%%%%%%%%%%%%%%%%%%%%%%%%%%%%%%%%%%%%%%%%%%%%%%%%%%
%%%%%%%%%%%%%%%%%%%%%%%%%%%%%%%%%%%%		Sixth-Turn Space
%%%%%%%%%%%%%%%%%%%%%%%%%%%%%%%%%%%%%%%%%%
%%%%%%%%%%%%%%%%%%%%%%%%%%%%%%%%%%%%%%%%%%%%%%%%%%%%%%%%%%%%%%%%%%%%%%%%%%%%%%%%%%%%%%%%%%%%%%%%%%%%%%%%%%%%%%
\subsection{Sixth-Turn Space ($E_{5}$)}
The hexagonal prism could be rotated by $60^{\circ}$ when stacked in
the $x$-direction.  The Fundamental Polyhedron's depth is now $6L_x$.
Once again, this is an orientable spacetime.  The interval is thus
\begin{align}
	\sigma = & \frac{1}{2} \Bigg\{ -(t - \tilde{t})^{2} + (x -
	\tilde{x} - n_{x} L_{x})^{2} \nonumber \\
	& \quad {}+ \left[ y - \cos \left( \frac{n_{x} \pi}{3} \right)
	\tilde{y} + \sin \left( \frac{n_{x} \pi}{3} \right) \tilde{z} -
	n_{y} L - \frac{1}{2} n_{z} L \right]^{2} \nonumber \\
	& \quad {}+ \left[ z - \sin \left( \frac{n_{x} \pi}{3} \right)
	\tilde{y} - \cos \left( \frac{n_{x} \pi}{3} \right) \tilde{z} -
	\frac{\sqrt{3}}{2} n_{z} L \right]^{2} \Bigg\}.
\end{align}
Figure~\ref{plot:E4E5} (b) shows our numerical result.

%%%%%%%%%%%%%%%%%%%%%%%%%%%%%%%%%%%%%%%%%%%%%%%%%%%%%%%%%%%%%%%%%%%%%%%%%%%%%%%%%%%%%%%%%%%%%%%%%%%%%%%%%%%%%%
%%%%%%%%%%%%%%%%%%%%%%%%%%%%%%%%%%%%	Hantzsche-Wendt Space
%%%%%%%%%%%%%%%%%%%%%%%%%%%%%%%%%%%%%%%%%%
%%%%%%%%%%%%%%%%%%%%%%%%%%%%%%%%%%%%%%%%%%%%%%%%%%%%%%%%%%%%%%%%%%%%%%%%%%%%%%%%%%%%%%%%%%%%%%%%%%%%%%%%%%%%%%
\section{Hantzsche-Wendt Space ($E_{6}$)}
To generate the Hantzsche-Wendt manifold, imagine a rectangular box of
sides $L_{x}$, $L_{y}$, and $L_{z}$ and three orthogonal,
nonintersecting axes, each on a face of the box.  When the box is
translated along one of the axes by its length, a half-turn corkscrew
motion about that axis is introduced \cite{Levin}.  The Fundamental
Polyhedron of the Hantzsche-Wendt Space is a rhombic dodecahedron
which circumscribes the rectangular box.  The Hantzsche-Wendt Space is
an orientable manifold.  The interval in this space is given by
\begin{align}
	\sigma = & \frac{1}{2} \Bigg\{ -(t - \tilde{t})^{2} + \left[ x -
	(-1)^{n_{b}} (-1)^{n_{c}} \tilde{x} - (n_{a} + n_{b})
	\frac{L_{x}}{2} \right]^{2} \nonumber \\
	& \quad {}+ \left[ y - (-1)^{n_{a}} (-1)^{n_{c}} \tilde{y} -
	(n_{b} + n_{c}) \frac{L_{y}}{2} \right]^{2} \nonumber \\
	& \quad {}+ \left[ z - (-1)^{n_{a}} (-1)^{n_{b}} \tilde{z} -
	(n_{a} + n_{c}) \frac{L_{z}}{2} \right]^{2} \Bigg\},
\end{align}
where $n_{a}$, $n_{b}$, and $n_{c}$ are any integers.

\begin{figure*}
	\centering
	\mbox{ \subfigure[$z =
	0$]{\scalebox{0.5}{\epsfig{figure=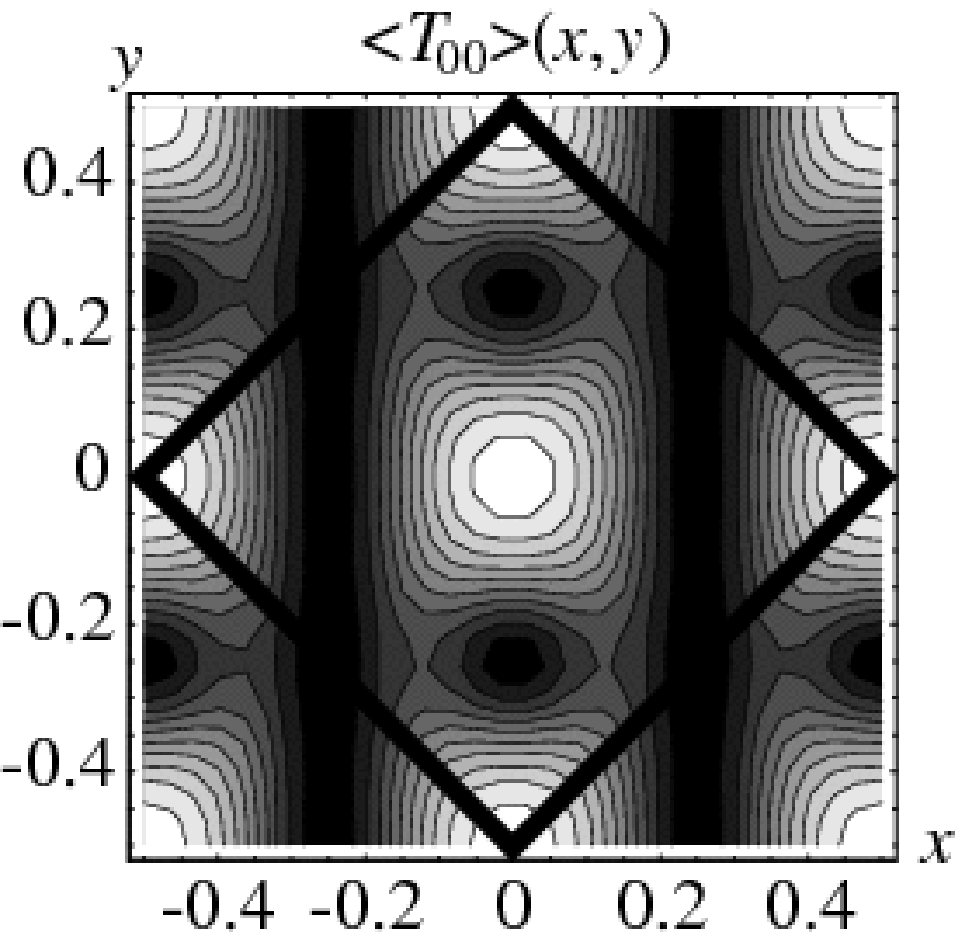}}} \quad
	\subfigure[$z =
	0.125$]{\scalebox{0.5}{\epsfig{figure=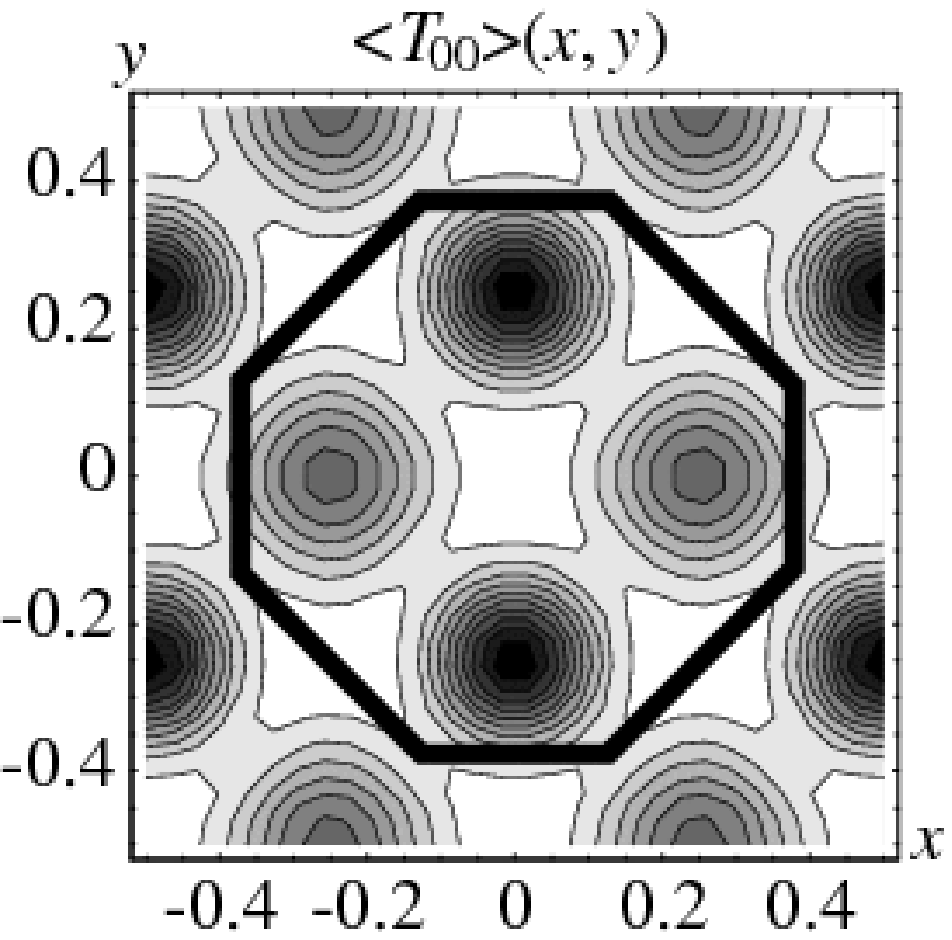}}} \quad
	\subfigure[$z =
	0.25$]{\scalebox{0.5}{\epsfig{figure=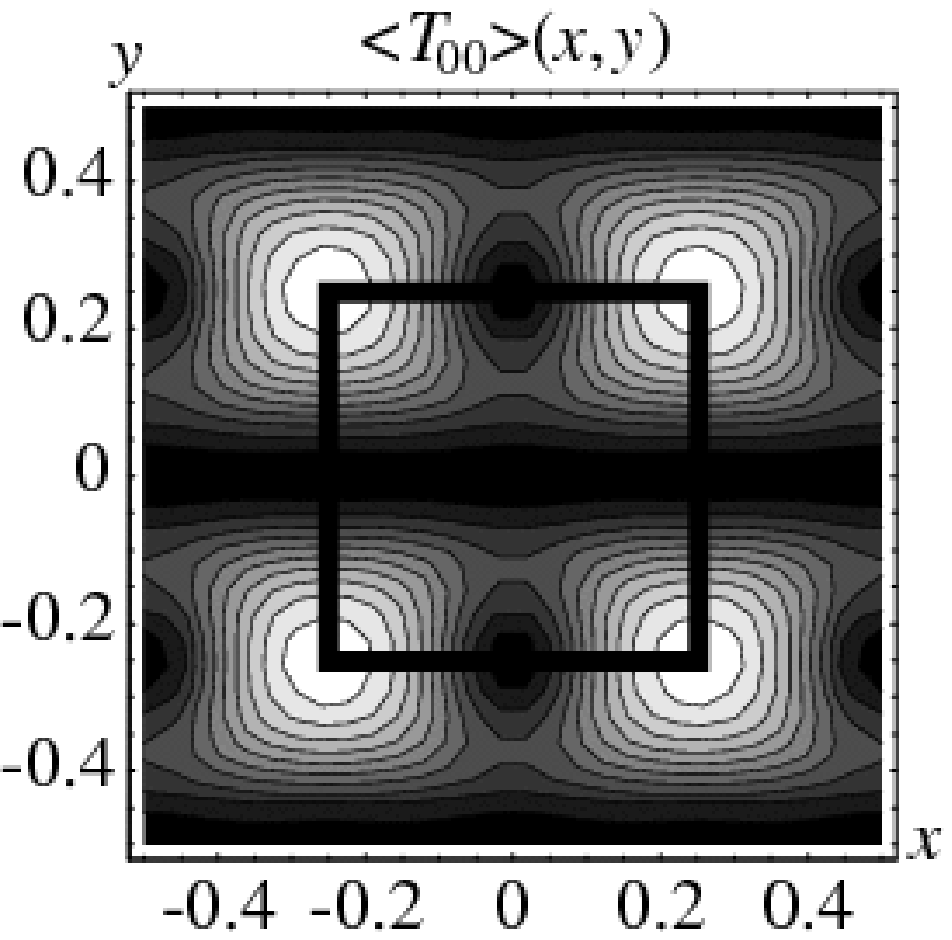}}} }
	\caption{Energy density in Hantzsche-Wendt Space.  Shown are the
	three planes in the Fundamental Polyhedron, indicated by the heavy
	black line within each plot.}
	\label{plot:E6}
\end{figure*}

The resulting energy density is a function of all three spatial
dimensions.  Shown in Figure~\ref{plot:E6} is slices of this space in
the $x$-$y$ plane at $z=0$, $0.125$, and $0.25$.  This is identical to
slices through the center of the Fundamental Polyhedron for the other
two planes, as well.  $L_{x}$, $L_{y}$, and $L_{z}$ have been set
equal to 1.

\section{Three Closed Dimensions -- Klein Bottle Cross-Section}
\label{sec:3-TorusKlein}
To construct the Klein Spaces, start with a flat rectangular box of
length $L_{x}$, width $L_{y}$, and height $L_{z}$.  Glue the faces in
the $y$- and $z$-directions in the usual way, but glue the faces in
the $x$-direction with a flip in the $y$-direction.  This flipping
results in a glide reflection in the $y$-coordinate under translations
in the $x$-direction.  The resulting space is a Klein bottle cross a
circle ($K^{2} \times S^{1}$).  While stacking the rectangles in the
$z$-direction, a vertical flip (in the $x$-coordinate), horizontal
flip (in the $y$-coordinate), or a half-turn (about the $z$-axis)
could be introduced, yielding a total of four Klein Spaces.  Because
of the flipping motions, all of the Klein Spaces are non-orientable.

%%%%%%%%%%%%%%%%%%%%%%%%%%%%%%%%%%%%%%%%%%%%%%%%%%%%%%%%%%%%%%%%%%%%%%%%%%%%%%%%%%%%%%%%%%%%%%%%%%%%%%%%%%%%%%
%%%%%%%%%%%%%%%%%%%%%%%%%%%%%%%%%%%% Klein Space
%%%%%%%%%%%%%%%%%%%%%%%%%%%%%%%%%%%%%%%%%%
%%%%%%%%%%%%%%%%%%%%%%%%%%%%%%%%%%%%%%%%%%%%%%%%%%%%%%%%%%%%%%%%%%%%%%%%%%%%%%%%%%%%%%%%%%%%%%%%%%%%%%%%%%%%%%
\subsection{Klein Space ($E_{7}$)}
Klein spaces have many possible Fundamental Polyhedra.  For
simplicity, we will choose a rectangular box of length $2L_{y}$, width
$L_{x}/2$ and height $L_{z}$.  Then, the interval in the basic Klein
Space is given by
\begin{equation}
	\sigma = \frac{1}{2} \{ -(t - \tilde{t})^{2} + (x - \tilde{x} -
	n_{x} L_{x}/2)^{2} + [y - (-1)^{n_{x}} \tilde{y} - 2n_{y}
	L_{y}]^{2} + (z - \tilde{z} - n_{z} L_{z})^{2} \}.
\end{equation}

\begin{figure*}
	\centering
	\epsfig{figure=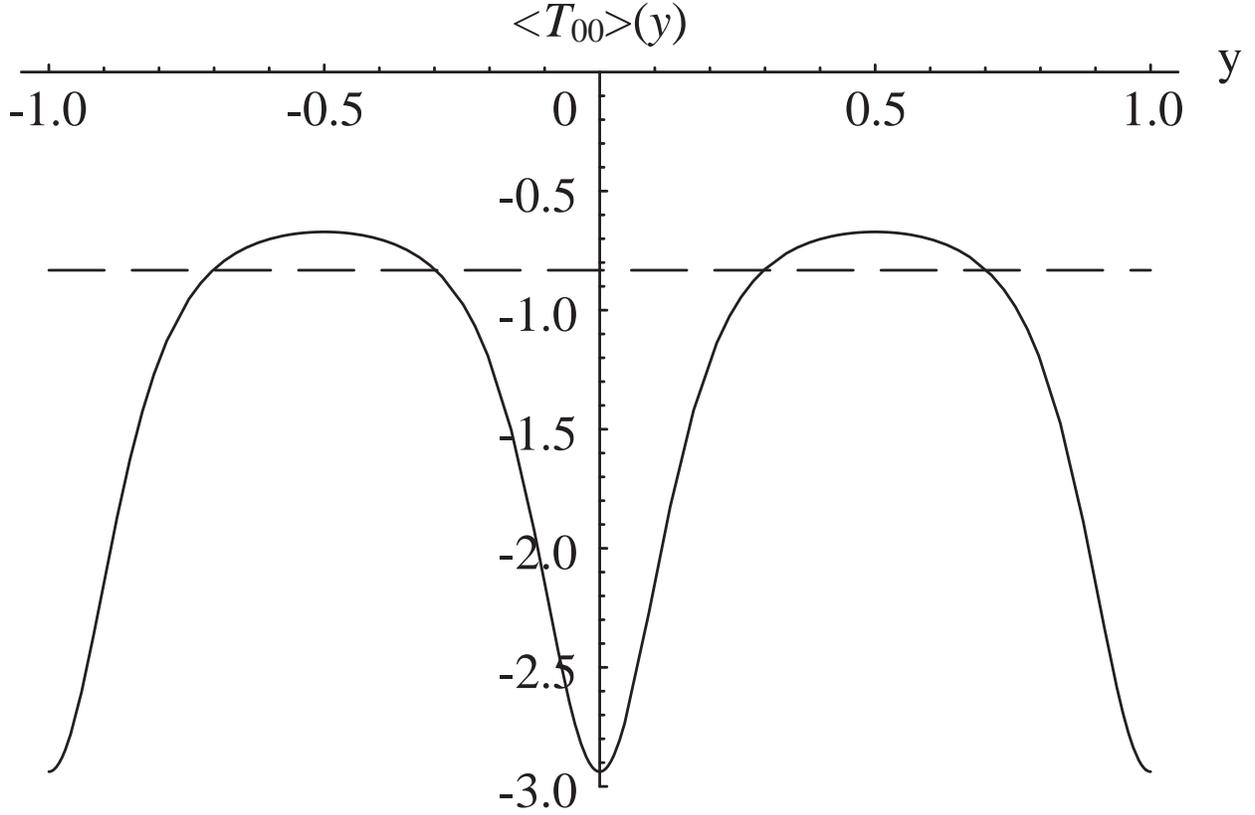,width=\textwidth}
	\caption{Comparing the vacuum energy densities in 3-Torus Space
	(dashed line) and Klein Space (solid lines), with $L_x = L_y = L_z
	= 1$.  The region is a single $z=const$ and $x=0$ slice of the
	Fundamental Polyhedra.  Note that we are showing two Fundamental
	Polyhedron of the 3-Torus Space, to aid in comparison.}
	\label{plot:compareE7E1}
\end{figure*}

The vacuum stress-energy tensor can be split into two summations over 
even and odd indices:
\begin{equation}
	\langle T_{\mu \nu} \rangle_{E_{7}} =
	\sideset{}{'}\sum_{\substack{n_{x}\ even \\ n_{y},\ n_{z}}}
	[\text{Klein}] + \sum_{\substack{n_{x}\ odd \\
	n_{y},\ n_{z}}} [\text{K-Flip}(y)],
\end{equation}
where
\begin{align*}
	[\text{Klein}] = & \frac{8}{\pi^2} \frac{1}{({L_x}^2 {n_x}^2 + 16
	{L_y}^2 {n_y}^2 + 4 {L_z}^2 {n_z}^2)^3 } \\
	& \quad \times \left( {L_x}^2 {n_x}^2 \diag\left[-1, -3, 1, 1
	\right] + 16 {L_y}^2 {n_y}^2 \diag\left[-1, 1, -3, 1 \right]
	\right.  \\
	& \quad \left.  + 4 {L_z}^2 {n_z}^2 \diag\left[-1, 1, 1, -3
	\right] \right)
\end{align*}
and
\begin{align*} 
	[\text{K-Flip} (y)] = &\frac{32}{3 \pi^2} \frac{1}{[{L_x}^2
	{n_x}^2 + 16(y - L_y n_y)^2 + 4 {L_z}^2 {n_z}^2 ]^3} \\
	& \quad \times \left( {L_x}^2 {n_x}^2 \diag\left[-1, -2, 0, 1
	\right] + 4 {L_z}^2 {n_z}^2 \diag\left[-1, 1, 0, -2 \right]
	\right).
\end{align*}
The even summation over $n_{x}$ produces a constant shift in the
vacuum energy density similar to the 3-Torus case whereas the odd
summation gives us a periodic $y$-dependence (see
Figure~\ref{plot:compareE7E1}).

%%%%%%%%%%%%%%%%%%%%%%%%%%%%%%%%%%%%%%%%%%%%%%%%%%%%%%%%%%%%%%%%%%%%%%%%%%%%%%%%%%%%%%%%%%%%%%%%%%%%%%%%%%%%%%
%%%%%%%%%%%%%%%%%%%%%%%%%%%% Klein Space with Horizontal Flip
%%%%%%%%%%%%%%%%%%%%%%%%%%%%%%%%%%%%%%%%%%
%%%%%%%%%%%%%%%%%%%%%%%%%%%%%%%%%%%%%%%%%%%%%%%%%%%%%%%%%%%%%%%%%%%%%%%%%%%%%%%%%%%%%%%%%%%%%%%%%%%%%%%%%%%%%%
\subsection{Klein Space with Horizontal Flip ($E_{8}$)}
When stacking rectangular Fundamental Polyhedra in the $z$-direction,
a horizontal flip in the $y$-direction can be introduced, yielding an
interval of
\begin{equation}
	\sigma = \frac{1}{2} \{ -(t - \tilde{t})^{2} + (x - \tilde{x} -
	n_{x} L_{x}/2)^{2} + [y - (-1)^{n_{x}} (-1)^{n_{z}} \tilde{y} -
	2n_{y} L_{y}]^{2} + (z - \tilde{z} - n_{z} L_{z})^{2} \}.
\end{equation}
By carefully examining this interval, and arranging the summations
correctly, we can get the exact same result as for the simple Klein
space.  We may write
\begin{equation}
	\langle T_{\mu \nu} \rangle_{E_{8}} =
	\sideset{}{'}\sum_{\substack{n_x\ and\ n_z\ even \\ or \\
	n_x\ and\ n_z\ odd \\ n_{y}}} [\text{Klein}] +
	\sum_{\substack{n_x\ even,\ n_z\ odd \\ or \\ n_x\ odd,\ n_z\ even
	\\ n_{y}}} [\text{K-Flip}(y)].
\end{equation}

%%%%%%%%%%%%%%%%%%%%%%%%%%%%%%%%%%%%%%%%%%%%%%%%%%%%%%%%%%%%%%%%%%%%%%%%%%%%%%%%%%%%%%%%%%%%%%%%%%%%%%%%%%%%%%
%%%%%%%%%%%%%%%%%%%%%%%%%%%% Klein Space with Vertical Flip
%%%%%%%%%%%%%%%%%%%%%%%%%%%%%%%%%%%%%%%%%%
%%%%%%%%%%%%%%%%%%%%%%%%%%%%%%%%%%%%%%%%%%%%%%%%%%%%%%%%%%%%%%%%%%%%%%%%%%%%%%%%%%%%%%%%%%%%%%%%%%%%%%%%%%%%%%
\subsection{Klein Space with Vertical Flip ($E_{9}$)}
Alternatively, we may flip in the vertical direction, $x$, as we
translate in the $z$-direction:
\begin{align}
	\sigma = & \frac{1}{2} \{ -(t - \tilde{t})^{2} + [x - (-1)^{n_{z}}
	\tilde{x} - n_{x} L_{x}/2]^{2} + [y - (-1)^{n_{x}} \tilde{y} -
	2n_{y} L_{y}]^{2} \nonumber \\
	& \quad {}+ (z - \tilde{z} - n_{z} L_{z})^{2} \}.
\end{align}
Each of four possible combinations of $n_x$ and $n_z$ gives a
different functional form:
\begin{align}
	\langle T_{\mu \nu} \rangle_{E_{9}} = &
	\sideset{}{'}\sum_{\substack{n_{x}\ even \\ n_{y} \\
	n_{z}\ even}} [\text{Klein}] + \sum_{\substack{n_{x}\ odd \\ n_{y}
	\\ n_{z}\ even}} [\text{K-Flip}(y)] \nonumber \\
	& \quad {}+ \sum_{\substack{n_x\ even \\
	n_{y} \\ n_{z}\ odd}} [\text{K-V-Flip}(x)] \sum_{\substack{n_x\
	odd \\ n_{y} \\ n_{z}\ odd}} [\text{K-Turn}(x,y)],
\end{align}
with
\begin{align*}
	[\text{K-V-Flip} (x)] = & \frac{128}{3 \pi^2} \frac{1}{[(4x - L_x
	n_x)^2 + 16 {L_y}^2 {n_y}^2 + 4 {L_z}^2 {n_z}^2]^3} \\
	& \quad \times \left( {L_x}^2 {n_x}^2 \diag\left[-1, 0, 1, -2
	\right] + 4 {L_y}^2 {n_y}^2 \diag\left[-1,0, -2, 1 \right] \right)
\end{align*}
and
\begin{align*}
	[\text{K-Turn} (x,y)] = & \frac{8}{3 \pi^2} \frac{1}{[(4x - L_x
	n_x)^2 + 16(y - L_y n_y)^2 + 4 {L_z}^2 {n_z}^2]^3} \\
	& \quad \times \Big\{ (4x - {L_x} {n_x})^2 \diag\left[-1, 1, -3,
	1 \right] \\
	& \quad + 16(y - {L_y} {n_y})^2 \diag\left[-1, -3, 1, 1 \right] +
	4{L_z}^2 {n_z}^2 \diag\left[-5, 1, 1, -7\right] \\
	& \quad {}+ 16 (4x - L_{x} n_{x}) (y - L_{y} n_{y})
	\left(\delta^{x}_{\ \mu} \delta^{y}_{\ \nu} + \delta^{y}_{\ \mu}
	\delta^{x}_{\ \nu}\right) \Big\}.
\end{align*}

%%%%%%%%%%%%%%%%%%%%%%%%%%%%%%%%%%%%%%%%%%%%%%%%%%%%%%%%%%%%%%%%%%%%%%%%%%%%%%%%%%%%%%%%%%%%%%%%%%%%%%%%%%%%%%
%%%%%%%%%%%%%%%%%%%%%%%%%%%% Klein Space with Half-Turn
%%%%%%%%%%%%%%%%%%%%%%%%%%%%%%%%%%%%%%%%%%
%%%%%%%%%%%%%%%%%%%%%%%%%%%%%%%%%%%%%%%%%%%%%%%%%%%%%%%%%%%%%%%%%%%%%%%%%%%%%%%%%%%%%%%%%%%%%%%%%%%%%%%%%%%%%%
\subsection{Klein Space with Half-Turn ($E_{10}$)}
The last space is the Klein Space with Half-Turn in which a corkscrew
motion of $180^{\circ}$ about the $z$-axis is introduced, generating
an interval of
\begin{align}
	\sigma = & \frac{1}{2} \{ -(t - \tilde{t})^{2} + [x - (-1)^{n_{z}}
	\tilde{x} - n_{x} L_{x}/2]^{2} + [y - (-1)^{n_{x}} (-1)^{n_{z}}
	\tilde{y} - 2n_{y} L_{y}]^{2} \nonumber \\
	& \quad {}+ (z - \tilde{z} - n_{z} L_{z})^{2}
	\}.
\end{align}
Interestingly, $\langle T_{\mu\nu} \rangle$ contains the same four
summations as those for the Klein Space with Vertical Flip except that
the indices for the last two summations are exchanged.
\begin{align}
	\langle T_{\mu \nu} \rangle_{E_{10}} = &
	\sideset{}{'}\sum_{\substack{n_{x}\ even \\ n_{y} \\
	n_{z}\ even}} [\text{Klein}] + \sum_{\substack{n_{x}\ odd \\ n_{y}
	\\
	n_{z}\ even}} [\text{K-Flip}(y)]  \nonumber \\
	& \quad {} + \sum_{\substack{n_{x}\ odd \\ n_{y} \\
	n_{z}\ odd}} [\text{K-V-Flip}(x)] + \sum_{\substack{n_{x}\ even \\
	n_{y} \\
	n_{z}\ odd}} [\text{K-Turn}(x,y)].
\end{align}

The vacuum energy densities in Klein Spaces with Horizontal Flip,
Vertical Flip, and Half-Turn deviate very little from that in the
simple Klein Space.  Flips and turns introduce only small modulations
in the vacuum energy density.

\section{Topological Effects on the Vacuum Energy Density}
\label{sec:Effects}
Despite many differences among the results described above, there are
quite a few similarities.  Every vacuum energy density calculated is
negative relative to the value in the simply-connected Minkowski
space.  All of the results are dependent on the size of the
Fundamental Polyhedron.  Specifically, the stress-energies of all the
spaces are proportional to the inverse-fourth power of the length of
the Fundamental Polyhedron.  Thus, as the size of the Fundamental
Polyhedra increases the vacuum energy density approaches zero.  This
is because the scalar field is allowed to have more modes, pushing the
value of vacuum density upward to zero.

By closing a single dimension, the vacuum energy density shifts
downward by a constant.  As we move to the 2-Torus and 3-Torus, the
energy density shifts even further.  Closing more dimensions reduces
the number of modes available to the quantum field (as noted by DeWitt
and others \cite{DeWitt}).  The wavelength must have discrete values
not only in the $x$-direction but also in the $y$- and $z$-directions.
This shift in the energy density is always a constant throughout the
universe and never position-dependent.

The vacuum energy density also reflects the symmetries of the
underlying Fundamental Polyhedron.  The densities of 1/4-Turn and
1/2-Turn spaces are repeating patterns of rectangular grids, like
squares on a checkerboard.  However, the densities of 1/6-Turn and
1/3-Turn spaces are repeating hexagon patterns, mirroring how the
Fundamental Polyhedron is tiled.

Moreover, the value of the vacuum energy density is dependent on the
cross section of the Fundamental Polyhedron.  For the Klein Spaces,
the cross section of the Fundamental Polyhedron is a Klein bottle,
resulting in a lower energy density than the 3-Torus spaces with the
toroidal cross section.  Likewise, the hexagonal torus of Third- and
Sixth-Turn Spaces reduces the energy density below the Minkowski
level, but not to the extent of the Klein bottle cross section.
Table~\ref{tab:constants} summarizes shifts in the constant,
position-independent part of the energy density due to various cross
sections of the Fundamental Polyhedron.

\begin{table*}
	\centering
	\begin{tabular}{|c|c|} 
		\hline
		Tiling & Energy Density \\
		\hline 
		\hline
		1-Torus & $-0.11$ \\
		2-Torus & $-0.31$ \\
		Rectangular 3-Torus & $-0.83$ \\
		Hexagonal prism & $-0.99$ \\
		Klein space & $-2.39$ \\
                Hantzsche-Wendt Space & $-0.32$ \\
		\hline
	\end{tabular}
	\caption{Possible tilings of three-dimensional space and their
	associated shift in vacuum energy density relative to Euclidean
	space.  For these calculations, $L_{x} = L_{y} = L_{z} = L = 1$ in
	all cases, only the constant, position-independent terms are
	calculated.}
	\label{tab:constants}
\end{table*}

Positional dependence of the stress-energy tensor appears if we add
flipping or corkscrew motions when gluing faces of the Fundamental
Polyhedra.  When a dimension undergoes a reflection or turn, the
stress-energy tensor becomes dependent on that direction.  For
example, in the Third-Turn Space, we rotate the $y$- and
$z$-coordinates, so $\langle T_{\mu\nu} \rangle$ depends on $y$ and
$z$ but is still independent of $x$.  If the flipped dimension is
closed, as the universe is traversed in that direction, the energy
density well is encountered again and again at a constant interval.

If the flipped dimension is open, a single well appears in the energy
density, and translational symmetry is broken.  For example, consider
Figure~\ref{plot:E17}, which is the plot of the energy density in the
1-Torus with Flip.  The energy density reaches a minimum where $y =
0$.  The vacuum energy density will be symmetric about this plane.  If
we were traveling in this universe in the $y$ direction, we would
eventually encounter this location where the energy density is at a
minimum.  Figure~\ref{plot:CompareE11E13E14} compares effects of flips
in different directions in 2-Torus spaces.  In the 2-Torus with
Vertical Flip, the flip occurs in an open dimension ($z$) resulting in
a single well in the energy density at $z = 0$.  In the 2-Torus with
Horizontal Flip, we have periodic wells in the $y$-direction because
the flip occurs in a closed dimension ($y$).

Direct comparisons of vacuum energy density plots reveal many
interesting features.  For example, Figure~\ref{plot:compareE12E15}
shows the results of adding an additional flip to the 2-Torus with
Half-Turn.  While they have the same value in the center of the
Fundamental Polyhedron, the flipping causes the energy density to
reach a lower value on the boundary of the Fundamental Polyhedron.

\begin{figure}
	\epsfig{file=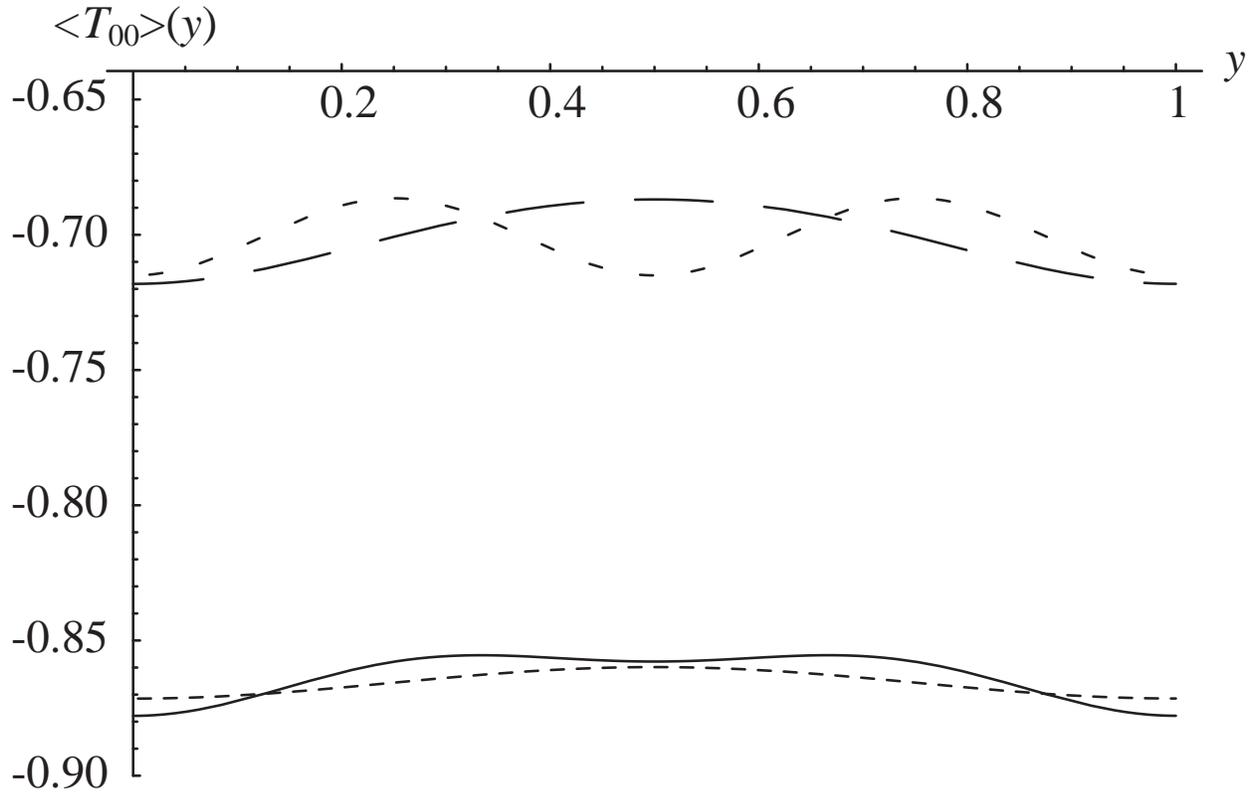,width=\textwidth}
	\caption{Comparison of 3-Torii with different angles of rotation.
	Shown are: Half-Turn Space (short dash, long gap), Quarter- Turn
	Space (long dash), Third-Turn Space (solid), and Sixth-Turn Space
	(short dash, short gap).  The region is a single $x=const$ and
	$z=0$ region of the Fundamental Polyhedron.  Note that for the
	Third- and Sixth-Turn Spaces, the energy density is shifted by one
	half-length of the Fundamental Polyhedron, to aid in comparison.}
	\label{plot:compareE2E3E4E5}
\end{figure}

Similarly, we notice the effects of various rotational angles by
examining Figure~\ref{plot:compareE2E3E4E5}.  This plot features the
corkscrew motions on the 3-Torii.  The strongest difference in energy
densities is due to switching from rectangular to hexagonal torus of
the Fundamental Polyhedron, while increased rotations (i.e. from
$90^{\circ}$ to $180^{\circ}$ rotations) do not change the average
value, but instead change the periodicity.

Our calculations in various Klein Spaces show that the amplitude of
the periodic wells in the vacuum energy density, introduced by
switching from the rectangular cross section to the Klein bottle cross
section, is much larger than the amplitude of the amplitude of the
modulations due to flips and turns.

\section{Conclusion}
\label{sec:Conclusion}
We have discovered that the vacuum expectation value of the
stress-energy tensor in non-trivial manifolds exhibits the following
qualitative behaviors : (1) increasing the number of closed dimensions
further lowers the vacuum energy density relative to the zero value of
the simply-connected Minkowski space, (2) the spatial pattern of the
vacuum energy density follows the shape of the chosen Fundamental
Polyhedron, (3) if a flip is added in a particular direction, then
there will be a positional dependence in that particular direction,
(4) if a turn is introduced, the stress-energy tensor will be
dependent on the coordinates in the plane perpendicular to the
rotational axis and will contain off-diagonal elements, and (5) the
chosen cross section of the Fundamental Polyhedron greatly effects the
relative vacuum energy densities, whereas flips and turns introduce
only a minor position-dependent shift.

\bibliography{sutter-tanaka}		
\bibliographystyle{apsrev}	
\nocite{*}

\end{document}